\documentclass[journal,12pt,onecolumn,draftclsnofoot,]{IEEEtran}

\usepackage{soul} % 导入 soul 包
\usepackage{color, xcolor}

\usepackage{cite}
\usepackage{algorithm}
\usepackage[noend]{algpseudocode}
\usepackage{amsmath,amssymb,amsfonts}
\usepackage{graphicx}
\usepackage{subfigure}
\usepackage{textcomp}
\usepackage{xcolor}
\usepackage{multirow}
\usepackage{makecell}
\usepackage{array}

\usepackage{diagbox}
    
\usepackage{pifont}

\usepackage{color}

\usepackage{hyperref}
  
\begin{document}
%\newcommand{\topic}[1]{\color{red}\textbf{#1}\color{black}}

%\title{Semantic Communication Systems for Speech Transmission}
%\title{Semantic Communication Systems for Speech Transmission with multi scale semantic alignment}
\title{Semantic-preserved Communication System for Highly Efficient Speech Transmission} 

%\author{XXX,~\IEEEmembership{Student Member,~IEEE,} and XXX,~\IEEEmembership{Member,~IEEE} 
\author{Tianxiao Han, Qianqian Yang\IEEEauthorrefmark{2}, Zhiguo Shi, Shibo He, Zhaoyang Zhang%
\thanks{This work will be presented in part at the IEEE Int'l Conf. on Commun.(ICC), Workshop on Semantic Commun., Seoul, South Korea, May 2022.}
\thanks{Tianxiao Han, Qianqian Yang, Zhiguo Shi and Zhaoyang Zhang are with the College of Information Science and Electronic Engineering, Zhejiang University, Hangzhou 310027, China (e-mail: {txhan,qianqianyang20\IEEEauthorrefmark{2},shizg,ning\_ming}@zju.edu.cn). Shibo He is with the Department of Control Science and Engineering, Zhejiang University, Hangzhou 310027, China (e-mail: s18he @zju.edu.cn).}% <-this % stops a space

\thanks{{\thefootnote}{*}This work is partly supported by the SUTD-ZJU IDEA Grant (SUTD-ZJU (VP) 202102), and partly by the Fundamental Research Funds for the Central Universities under Grant 2021FZZX001-20.}
}

\maketitle

\begin{abstract}
Deep learning (DL) based semantic communication methods have been explored for the efficient transmission of images, text, and speech in recent years. In contrast to traditional wireless communication methods that focus on the transmission of abstract symbols, semantic communication approaches attempt to achieve better transmission efficiency by only sending the semantic-related information of the source data. In this paper, we consider semantic-oriented speech transmission which transmits only the semantic-relevant information over the channel for the speech recognition task, and a compact additional set of semantic-irrelevant information for the speech reconstruction task. We propose a novel end-to-end DL-based transceiver which extracts and encodes the semantic information from the input speech spectrums at the transmitter and outputs the corresponding transcriptions from the decoded semantic information at the receiver. In particular, we employ a soft alignment module and a redundancy removal module to extract only the text-related semantic features while dropping semantically redundant content, greatly reducing the amount of semantic redundancy compared to existing methods. We also propose a semantic correction module to further correct the predicted transcription with semantic knowledge by leveraging a pretrained language model. For the speech to speech transmission, we further include a CTC alignment module that extracts a small number of additional semantic-irrelevant but speech-related information, such as duration, pitch, power and speaker identification of the speech for the better reconstruction of the original speech signals at the receiver. We also introduce a two-stage training scheme which speeds up the training of the proposed DL model. The simulation results confirm that our proposed method outperforms current methods in terms of the accuracy of the predicted text for the speech to text transmission and the quality of the recovered speech signals for the speech to speech transmission, and significantly improves transmission efficiency. More specifically, the proposed method only sends 16\% of the amount of the transmitted symbols required by the existing methods while achieving about 10\% reduction in WER for the speech to text transmission. For the speech to speech transmission, it results in an even more remarkable improvement in terms of transmission efficiency with only 0.2\% of the amount of the transmitted symbols required by the existing method while preserving comparable quality of the reconstructed speech signals. 
\end{abstract}

\IEEEpeerreviewmaketitle

\section{Introduction}
The continuously increasing demand for communication causes the explosion of wireless data traffic, and places a heavy burden on the current infrastructure of communication systems. Semantic communication is a promising technology for next generation communications because of its great potential of significantly improving transmission efficiency\cite{qin2021semantic_survey}. Unlike traditional communication systems, which focus on transmitting symbols while ignoring semantic content, semantic communication focuses on gathering semantic information from the source and recovering the same semantic information at the receiver. Therefore, concentrated semantic information will be transmitted to the receiver instead of directly mapped bit sequences from the source. By doing so, to transmit the same amount of information, the required resources for semantic communication will be reduced significantly. Moreover, semantic communication has been proved to be more robust than traditional communication systems\cite{qin1}, especially in harsh channel conditions.
% # 语义通信的发展历程和现状
%The idea of semantic communication has been proposed by Weaver at the beginning of modern communication\cite{shannon1949mathematical}. Three levels of communication was proposed then. The first level is "How accurately can symbols of communication be transmitted". The second level is "how precisely do the transmitted symbols convey the desired meaning". And the third level is "How effectively does the received meaning affect conduct in the desired way".

The idea of semantic communication has been proposed by Weaver at the beginning of modern communication\cite{shannon1949mathematical}. Following this preliminary work, Carnap and Bar-hillel \cite{carnap1952outline} gave an information theoretic definition of semantic information, which uses the logical probability of a sentence to calculate the semantic entropy that measures the amount of semantic information contained in the sentence. Another more recent follow-up work by Floridi \cite{floridi2004outline} proposed to use the truth likeness of a sentence instead of logical probability to quantify the amount of the semantic information. Semantic-aware data compression has been investigated in \cite{bao2011towards} to design a general transceiver system which encodes and decodes the semantic information at the transmitter and the receiver, respectively. It has been further investigated in \cite{basu2014preserving} by leveraging a shared knowledge base between the transmitter and the receiver to improve compression efficiency. However, before the boom of deep learning, there has not been an effective way to actually perform semantic communication of content.

With the emergence of deep learning techniques on image processing and language processing, there have been several works on semantic communications which show the superiority over the traditional methods.
%image
A CNN model has been presented in \cite{bourtsoulatze2019deep} to enable joint source and channel coding (JSCC) for wireless image transmission, which can recover images under limited bandwidth and low SNR conditions, and achieves efficient image transmission. In \cite{kurka2021bandwidth}, the authors designed a layered wireless image transmission scheme with multiple refinement layers of different compression rates, which can adapt the reconstruction quality of the images according to the bandwidth and channel conditions. Similarly, the authors \cite{zhang2022wireless} proposed a  multi-level semantic-aware communication for image transmission, which showed the significance of high-level semantic information, such as the image captioning information. 
%text

The semantic communication system for text was first proposed in \cite{farsad2018deep}, where recurrent neural network (RNN) is used as encoder and decoder to extract the semantic information and recover texts from it. This work has been further extended in \cite{rao2018variable} which has developed a variable length joint source and channel coding scheme for text that dynamically encodes the input text to transmitted symbols of variable lengths. The authors in \cite{qin1} have also proposed an efficient and robust semantic-oriented transmission scheme, the deep learning model of which was then further compressed to be able to work on IoT devices \cite{xie2020lite} using the model quantization and pruning. 
%other
\cite{xie2021task} and \cite{xie2021task1} designed semantic communication systems that are capable of multimodal data transmission for tasks, such as visual question answering. For the wireless video transmission, a semantic system has been developed in \cite{tung2021deepwive} which exploits reinforcement learning to optimize bandwidth allocation and GoP sizes. For the task-oriented communication, the authors in \cite{shao2021learning} utilized the information bottleneck principle to find a compact representation for a specific task while preserving the semantic-relevant information.

%Many good works about joint source and channel coding follows this principle, and further explore the communication efficiency of specific tasks of machines and compression methods. 

For the transmission of speech, attention-based semantic communication has been developed to recover speech signals at the receiver\cite{Weng2101:Semantic}, which views each frame of the speech spectrums as an image, and utilizes convolutional neural network to compress the speech spectrum. A federal learning-based approach has been proposed in \cite{tong2021federated} to further improve the accuracy of recovered speech signals at the receiver. %The transmission of speech signal is a promising task, but as the experiments reveal, the result of these methods will introduce mistakes for the downstream task, especially the speech recognition task. That is to say, the method to transmit speech-related features will bring noise over the semantic information the speech signal carries. Instead, we can directly transmit the text-related features of speech, first obtain the corresponding text at the receiver, and then recover the speech from the received text with additional speech information extracted from the original speech.
A speech recognition semantic communication system has been developed in \cite{weng2021semantic}, which reconstructs text transcription of the speech signals at the receiver by transmitting text-related semantic features. However, the connectionist temporal classification (CTC) based approach proposed by \cite{graves2006connectionist} encodes each speech spectrum frame into the same amount of transmitted symbols, while ignoring the difference in semantic significance of each frame, which may degrade the transmission efficiency. 

% ## 本文是如何使用新方法来解决这个问题的，还有什么不一样的点，进步有多少大 再长一点
In this paper, we propose a novel semantic communication system for the transmission of speech for both the speech recognition and recovery tasks. To improve the transmission efficiency, we employ an attention-based alignment module to enforce the amount of the semantic features to be transmitted to be close to that of the corresponding text content. All the repeated and semantically irrelevant features are further dropped by a redundancy removal module. In this way, only the semantic-relevant information is extracted from the input speech spectrum and sent over the channel. To enhance the correctness of the predicted transcription at the receiver, we use a beam search decoder to find the most possible subwords sequences, and a semantic corrector to avoid semantic errors by leveraging a pretrained language model. For the successful recovery of the speech signal at the receiver, we also propose an additional speech information extractor that extracts a compact set of additional speech related information from the speech spectrums at the transmitter, which contains the duration, pitch and power information. The simulation results illustrate the remarkable improvement in the transmission efficiency of the proposed approach over the existing methods while boosting the accuracy of the predicted transcription and preserving the reconstruction quality of the speech signals at the same time. 

The main contributions of this paper can be summarized as follows:
\begin{itemize}
\item We propose a highly efficient semantic communication system for both the speech to text transmission and the speech to speech transmission, which, in particular, exploits an attention-based soft alignment module and a redundancy removal module, which extracts only the text-related semantic features and drops semantically irrelevant features to significantly improve the transmission efficiency.

\item For the successful prediction of the corresponding transcription, We employ a beam search semantic decoder together with a language model based semantic corrector to obtain the most possible transcription of the speech signals with semantic errors corrected by learnt semantic knowledge. Different from the existing methods that use characters or words as tokens, we use subwords instead, which preserves semantic meaning and avoids the problem of unseen words at the same time, thus improving the accuracy of the predicted transcription. 

\item For the successful recovery of the speech signal at the receiver, we also adopt a CTC alignment based additional speech information extractor that extracts a compact set of additional speech related information, including the duration, pitch and power information, from the speech spectrums at the transmitter, which are then utilized to synthesize the speech signals by a speech reconstructor module at the receiver.

\item  We propose a two-stage training method, which speeds up the training of the proposed model by training different parts at each stage. The numerical results validate the effectiveness and efficiency of the proposed method in both the text recognition performance and speech recovery. 
%We apply a two-stage training method, which trains the process of speech to text and text recovery separately with multi-task training methods. 
\end{itemize}

The rest of this article is organised as follows. Section \uppercase\expandafter{\romannumeral2} introduces the system model of the considered semantic communication problem and the performance metrics. Section \uppercase\expandafter{\romannumeral3} details the proposed deep learning-based approach semantic communication. Simulation results are presented in Section \uppercase\expandafter{\romannumeral4} and Section \uppercase\expandafter{\romannumeral5} concludes the paper.

\emph{Notation}: The boldface letters are used to represent vectors and matrixs, and single plain letters denote scalars. Given a vector $\boldsymbol x$, $x_i$ indicates its $i$th component, $\left\|\boldsymbol x\right\|$ denotes its Euclidean norm. Given a matrix $\boldsymbol Y$, $\boldsymbol Y\in\mathfrak R^{M\times N}$ indicates that $\boldsymbol Y$ is a matrix of size $M\times N$. $\mathcal{CN}(\boldsymbol m,\;\boldsymbol V)$ denotes multivariate circular complex Gaussian distribution with mean vector $\boldsymbol m$ and co-variance matrix $\boldsymbol V$.

\section{SYSTEM MODEL}
%% 本章介绍语音识别的语义通信是在干什么
In this section, we present the system model of the considered semantic communication system for highly efficient speech transmission, which aims to send compact semantic information of the input speech over the channel. This system is able to accomplish two different transmission tasks of speech signals, where one is to recover the corresponding text at the receiver, referred to as \textit{speech to text} transmission, and the other is to recover the speech signal, referred to as \textit{speech to speech} transmission. We also introduce the metrics to evaluate the performance of the proposed model for speech to text and speech to speech transmission.

\begin{figure}
\includegraphics[width=0.65\textwidth]{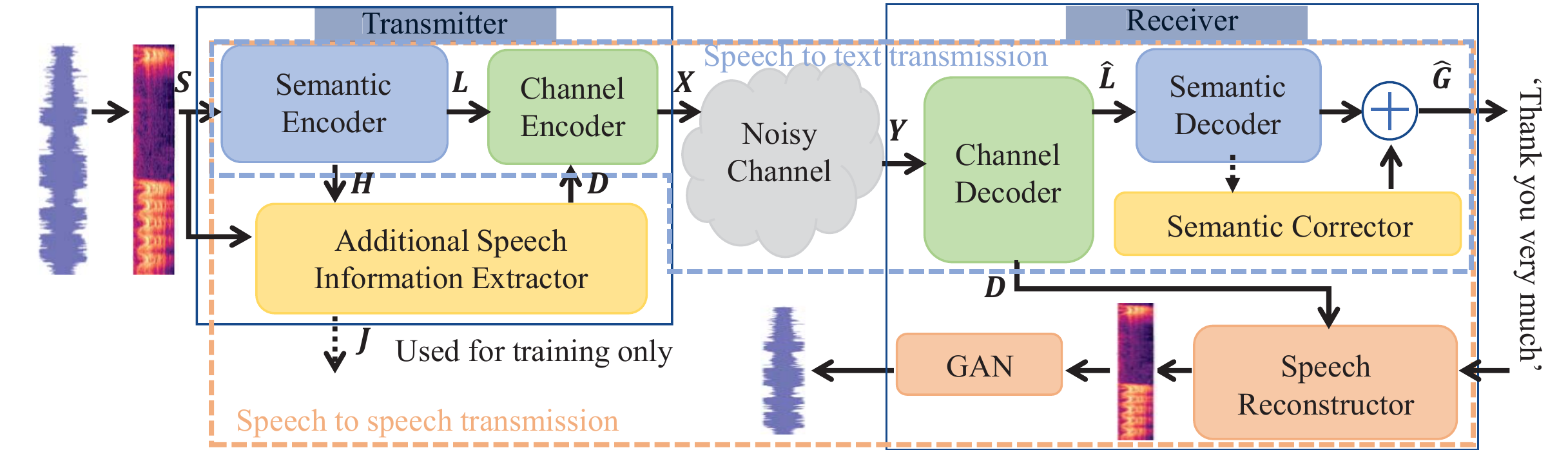}
\centering
\caption{The system model for speech transmission.}  % figure title
%\centering
\label{model}  
\end{figure}
    
\subsection{Speech Transmission System}

The considered semantic communication system for speech transmission consists of two parts as shown in Fig. \ref{model}: the transmitter and the receiver. The speech signal is sampled at 16 kHz, where we apply a 25 ms Hamming window and 10 ms shift. Then we compute fast Fourier transform(FFT) and filter banks(fbanks)\cite{hermansky1990perceptual } of each Hamming window to get the speech spectrum, which is then input to the transmitter, denoted by $\boldsymbol S=\left[{S_1,S_2,...,S_n }\right]$, where $n$ is the number of frames. For the \textit{speech to text} transmission, the receiver aims to output the corresponding transcription of the input speech, denoted by $\boldsymbol G=\left[{G_1,G_2,...,G_m}\right]$, where $G_i\in \mathcal{V}$ is the $i^{th}$ subword in the transcription, $m$ is the number of subwords in the transcription, and $\mathcal{V}$ denotes the vocabulary, which contains all the possible subwords in the speech. The mapping between $\boldsymbol S$ and $\boldsymbol G$ is called \textit{alignment} \cite{bahdanau2015neural}, which is a core task for speech recognition. For the \textit{speech to speech} transmission, the receiver aims to recover the speech spectrum $\widehat{\boldsymbol S}=\left[{\widehat{S}_1, \widehat{S}_2,..., \widehat{S}_n }\right]$ with the recovered text $\boldsymbol G$ and the additional speech related information received, which is then input to a generative adversarial network(GAN) to synthesize the speech. 

We note that subwords\cite{bazzi2002modelling} are used here as the tokens to construct the corresponding transcriptions of the input speech instead of characters or words. For example, subwords \textit{te, le, s, c, o, pe} constitute the word ``telescope''. Compared to using words as tokens as in our previous work \cite{han2022semantic}, it solves the problem of unknown or out-of-vocabulary (OOV) words since any new word can be tokenized into a set of subwords. It also has the advantage of preserving more semantic information than the methods that use characters as tokens in existing work\cite{weng2021semantic}. In this paper, we employ a pretrained subword model that has learnt a set of subword units from a text dataset\cite{kudo2018sentencepiece}.

The transmitter of the considered semantic communication system consists of a semantic encoder, an additional speech information extractor, and a channel encoder. The semantic encoder derives a compact latent semantic representation $\boldsymbol L$ and an intermediate semantic representation $\boldsymbol H$ from the input spectrum $\boldsymbol S$. Then the intermediate semantic representation $\boldsymbol H$ is input to the additional speech information extractor together with the input spectrum $\boldsymbol S$ to compute all the additional speech information, such as speech duration, pitch and power, denoted by $\boldsymbol D$. Then, the channel encoder maps $\boldsymbol L$ and $\boldsymbol D$ into symbols, denoted by $\boldsymbol X$, to be transmitted over physical channels. The received signals at the receiver are given by \begin{equation} {\boldsymbol Y=\boldsymbol h \ast \boldsymbol X+\boldsymbol w}, \label{channel}  \end{equation} where $\boldsymbol h$ represents the channel coefficients, and $\boldsymbol w\sim\mathcal{CN}(0,\;\sigma^2\mathbf I)$ denotes the independent and identically distributed (i.i.d.) complex Gaussian noise, where $\sigma^2$ is the noise variance, and $\mathbf I$ is the identity matrix.

The receiver consists of a channel decoder, a semantic decoder, a semantic corrector, a speech reconstructor and a pretrained GAN vocoder \cite{ren2020fastspeech}. The received signal, $\boldsymbol Y$, is first mapped back into the latent semantic representation $\widehat{\boldsymbol L}$ and additional speech information $\widehat{\boldsymbol D}$ by a channel decoder. $\widehat{\boldsymbol L}$ is then converted into the predicted transcriptions $\widehat{\boldsymbol G}$ by the semantic decoder with the help of a semantic corrector which keeps semantic information from previous context. For the speech to speech transmission, the predicted transcription $\widehat{\boldsymbol G}$ is then further processed to recover the speech spectrum by the semantic reconstructor with the received additional speech information $ \widehat{\boldsymbol D}$ including speech duration, pitch and power. Then a pretrained GAN vocoder recovers the desired speech signals from the reconstructed speech spectrum. We note here that in this considered communication system, only the meaning related information is transmitted even for the speech to speech transmission, with a compact set of additional speech related features. Experimental results, which will be presented in the sequel, reveal that this semantic-oriented transmission significantly improves the transmission efficiency while preserving the reconstruction qualities of either the corresponding transcription or the speech signals.

\subsection{Performance Metrics}
We employ the word-error-rate (WER\cite{klakow2002testing}) and the semantic similarity score \cite{qin1} as the performance metrics to evaluate the performance of the considered \textit{speech to text transmission}. WER is calculated by 
\begin{equation}
WER=\frac{S+D+I}{N},
\label{WER}
\end{equation}
where $S$, $D$, $I$ and $N$ denote the numbers of word substations, word deletions, word insertions, and the number of words in the transcription $\boldsymbol G$, respectively. 

The semantic similarity score that quantifies the sentence similarity between the predicted transcription $\widehat{\boldsymbol G}$, and  the original transcription, $\boldsymbol G$, is given by
\begin{equation}\label{similar}
   {\text{similarity}}\left( {\widehat{\boldsymbol G}},{\boldsymbol G} \right) =  \frac{{B \left( { \widehat{\boldsymbol G}} \right) \cdot { B {{\left(\boldsymbol {G} \right)}^T}}}}{{|| {B  {\left( \widehat{\boldsymbol G}\right)}} ||\cdot|| { B {\left(\boldsymbol {G} \right)}} ||}},
\end{equation}
where $ {B(\cdot) }$ represents sentence embedding by a pre-trained text embedding model, i.e., Bidirectional Encoder Representations from Transformers (BERT)\cite{devlin2018bert}. This sentence similarity score is a number between 0 and 1, which indicates how similar one sentence is to another, with 1 representing semantically equivalent and 0 representing not relevant at all.

For the \textit{speech to speech} transmission, We use the Mel cepstral distortion (MCD)\cite{kubichek1993mel} as the performance metrics to evaluate the reconstruction quality of the speech spectrum. The smaller the MCD is, the closer the reconstructed speech spectrum is to the original spectrum.
Recall that $S$ and $\widehat {S}$ denote the input spectrum and predicted spectrum, respectively. We align $S$ and $\widehat {S}$ using Dynamic Time Warping(DTW) algorithm, and calculate MCD by:
\begin{equation}
\operatorname{MCD}(S , \hat{S})=\frac{10}{\ln (10)} \sqrt{2 \sum_{t=1}^{T}\left\|S_{t}-\hat{S}_{t}\right\|},
\label{PESQ}
\end{equation}
where t denotes the index of the timestep, and T is the total number of timesteps.

In addition to the MCD, we also employ a  subjective quality evaluation test, that is, Mean opinion score (MOS)\cite{ribeiro2011crowdmos}, as another performance metric to further evaluate the quality of the recovered speech. The MOS rates the naturalness of the synthesized speech, which is obtained by having a group of raters, who are the native speakers of the same language, listening to the generated audio through the same headphones. The rates give scores in a five-point numeric scale, as shown in Table \ref{MOStable}, according to the naturalness of the speech samples, with 5 representing the highest perceived quality and 1 representing the lowest perceived quality. 
\begin{table}
\centering
\caption{Definition of the MOS five-point scale}
\label{MOStable}
\begin{tabular}{|c|c|c|}
\hline
Description & score \\
\hline
Excellent - Completely natural speech & 5 \\
\hline
Good - Mostly natural speech & 4 \\
\hline
Fair - Equally natural and unnatural speech & 3 \\
\hline
Poor - Mostly unnatural speech & 2 \\
\hline
Bad - Completely unnatural speech & 1 \\
\hline
\end{tabular}
\end{table}

\section{Proposed Semantic Communication for Speech Transmission}
\begin{figure*}[tbp]
\includegraphics[width=\textwidth]{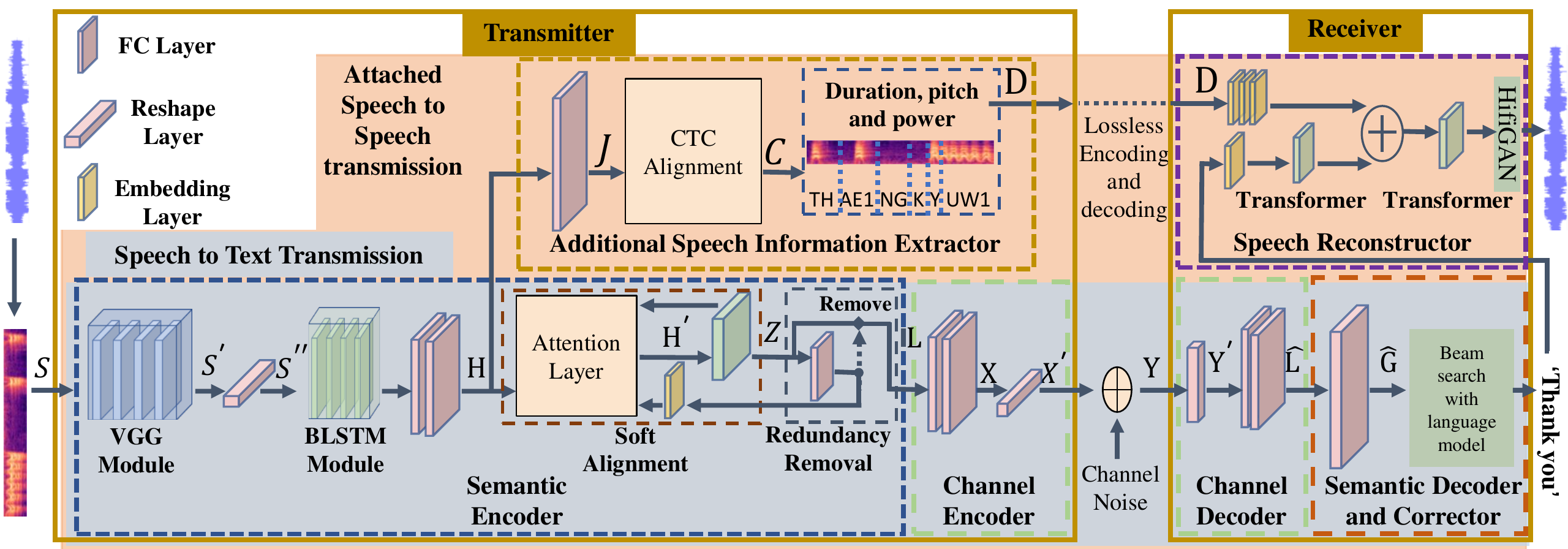} % image name
\centering % central position
\caption{The proposed system architecture for the speech transmission semantic communication system.} 
\label{dlmodel}  
\end{figure*}

The proposed semantic communication system contains two subsystems for speech transmission as shown in Fig. \ref{dlmodel}. The basic subsystem is for speech to text transmission, which extracts the semantic-related information from the speech spectrums and eliminates semantic-irrelevant redundancy in it. This semantic information is sent over the channel to the receiver, and then used to reconstruct the corresponding transcription of the original speech. The second subsystem, which is built upon the basic subsystem, aims for the speech to speech transmission. In addition to the semantic-related information, a set of semantic irrelevant information, which, however, helps the reconstruction of the speech signal at the receiver, is also extracted from the speech signals and sent to the receiver including the duration, pitch and power information of the speech. We describe the component that constitutes these two subsystems in detail in the sequel.

\subsection{Speech to Text Transmission}
As depicted in Fig. \ref{dlmodel}, the basis subsystem for the speech to text transmission consists of a semantic encoder, a channel encoder, a channel decoder, and a semantic decoder and corrector, where the semantic encoder extracts the semantic relevant information and removes the redundancy from the speech spectrums, and the channel encoder then processes this information into transmitted symbols. The received symbols at the receiver are then decoded by the channel decoder back to the semantic information, which is then used to reconstruct the transcriptions by the semantic decoder and corrector. We introduce each module in detail in the following.

\subsubsection{Semantic Encoder}
The semantic encoder has five components, including a VGG module \cite{vgg}, a bidirectional Long Short Term Memory (BLSTM) module, a FC module, a soft alignment module, and a redundancy removal module. We recall that the input speech spectrums, $\boldsymbol S\in\mathfrak R^{B\times N \times 40 }$ are acquired by applying 25 ms Hamming window and 10 ms shift on the speech signals, and then computing FFT to get 40 fbanks coefficients, where $B$ is the batch size, $N$ is the number of frames, and $40$ is the number of coefficients. The sequence of speech spectrums $\boldsymbol S$ is first fed into the VGG module to obtain a latent representation $\boldsymbol S^{\prime}\in\mathfrak R^{B\times \frac{N}{4} \times 10 \times 256}$, that is, 256 feature maps of size $\frac{N}{4} \times 10$ for each input. And the reshape layer concatenates all the 256 channels of $\boldsymbol S^{\prime}$ and reshapes the latent representation to $ \boldsymbol S^{\prime \prime}\in\mathfrak R^{B\times \frac{N}{4} \times 2560}$. Then, $\boldsymbol S^{\prime \prime}$ is input into the BLSTM module, which generates the intermediate features that preserve the temporal correlation in the input sequence. The followed FC module further compresses these intermediate features into a representation $\boldsymbol H \in\mathfrak R^{B\times \frac{N}{4} \times 1024}$, which is then fed into the soft alignment module. The soft alignment module then extracts the transcription-related information, and outputs latent semantic representations $\boldsymbol Z \in\mathfrak R^{B \times q \times 1024}$, which is further input to the redundancy removal module to remove the semantic-irrelevant redundancy and generate a more compact latent semantic representation ${\boldsymbol L \in\mathfrak R^{B \times c \times 1024}}$, where $c$ is the length of this representation after removing the redundant parts. 

\begin{figure}[tbp]
\includegraphics[width=0.65\textwidth]{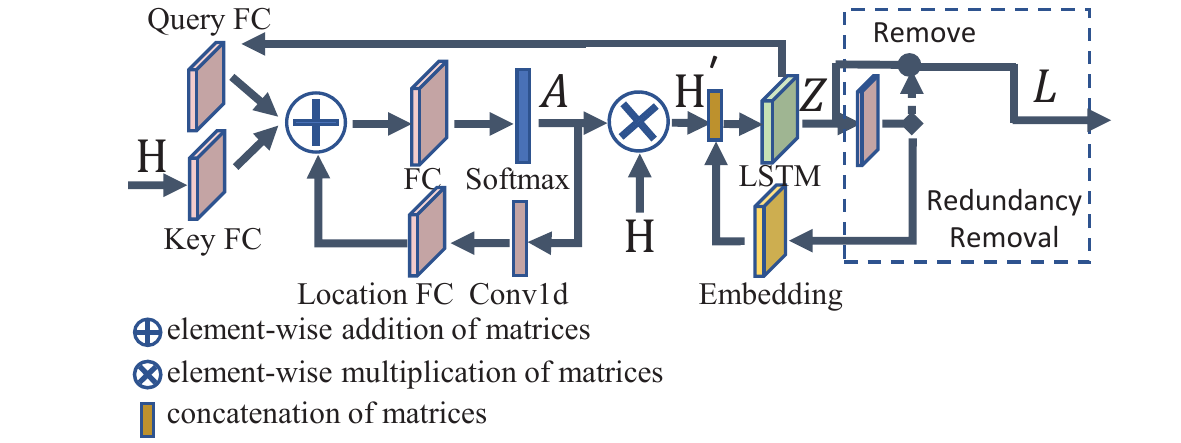} % image name
\centering % central position
\caption{The proposed soft alignment module with an attention module and a LSTM layer}  % figure title
\label{att}  
\end{figure}

\begin{figure}[tbp]
\includegraphics[width=0.5\textwidth]{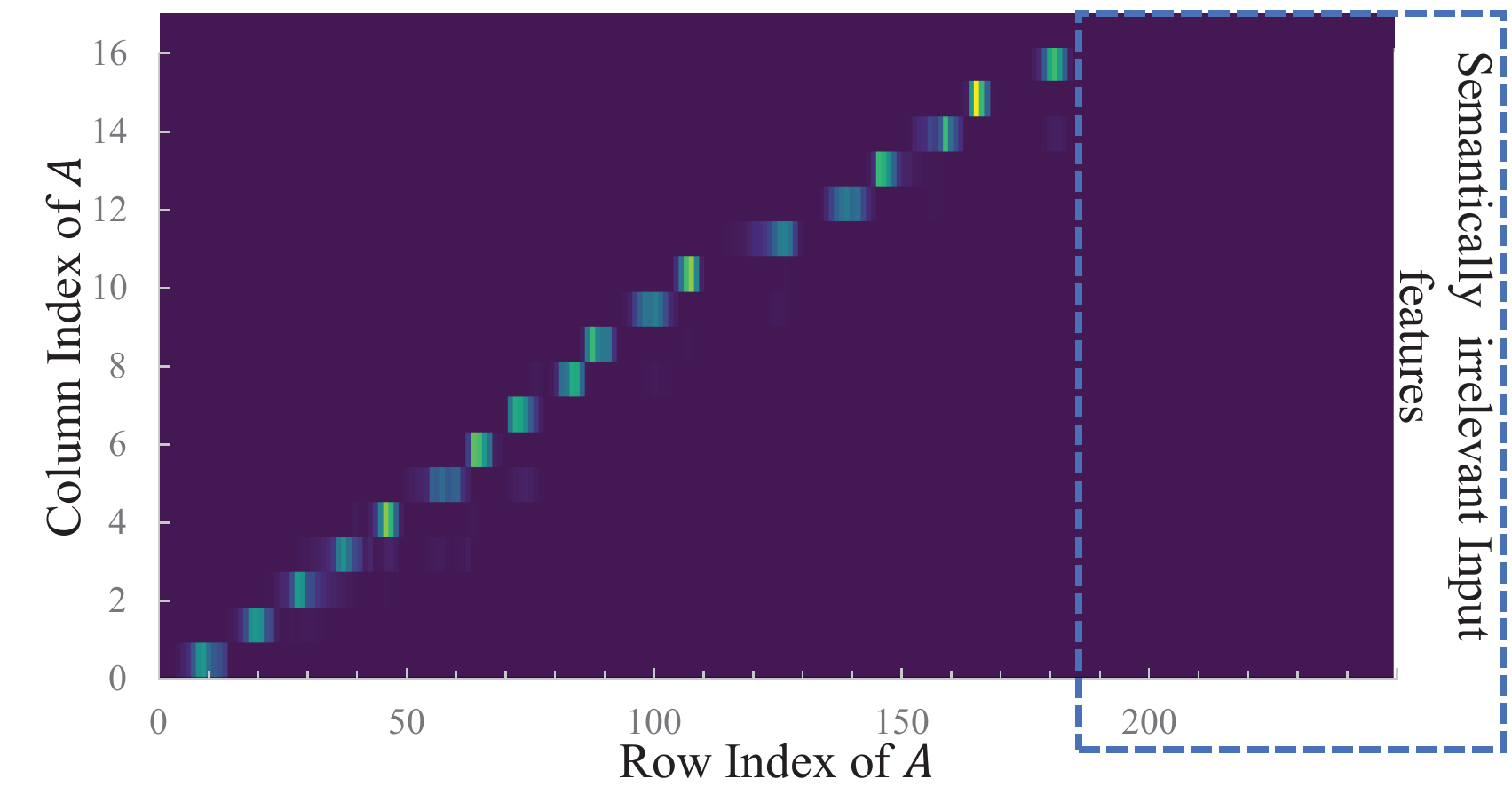} % image name
\centering % central position
\caption{Heatmap of attention scores generated by the attention mechanism. Larger values appear yellow, and lower values appear purple}  % figure title
\label{att_scores}  
\end{figure}

The soft alignment module uses an attention mechanism together with a LSTM layer to extract semantic features, as shown in Fig. \ref{att}.
The attention mechanism \cite{chorowski2015attention} is adopted for our soft alignment module to get the alignment of speech with its semantic text. In order to acquire the alignment, we need to compute the weight, also referred to as attention scores, assigned to each element of the input latent features, $\boldsymbol H$, by the query information and the corresponding key information. The query information is derived by passing the hidden states of the LSTM layer through a fully connected layer, and the key information is derived by another fully connected layer with $\boldsymbol H$ as the input. The query and key information are then combined with the feedback information through element-wise addition, as shown in Fig. \ref{att}. Then this combined information is fed into a FC layer and then a softmax layer to get the normalization scores $\boldsymbol A $, which feedback through a 1D convolution layer and a FC layer to derive the location information. Each value in the normalized attention scores $\boldsymbol A \in\mathfrak R^{B \times q \times \frac{N}{4}}$ multiplies its corresponding value in $\boldsymbol H$ to get the latent semantic representation, $\boldsymbol H^{\prime} \in\mathfrak R^{B \times q \times 512}$, where $q$ is the  length of latent semantic representation, which depends on the semantic information the signal is carrying. And $\boldsymbol H^{\prime}$, concatenated with the feedbacked embeddings, is fed into a LSTM layer, which outputs $\boldsymbol Z \in\mathfrak R^{B \times q \times 1024}$ as the input to the redundancy removal module. We present one example of the derived attention scores matrix $\boldsymbol A$ in the form of a heat map in Fig. \ref{att_scores}. We can observe that only a few elements of this matrix are with values that are not close to zero. As it is revealed by the numerical results, q is always smaller than 10 percent of N. So the proposed soft alignment module leads to better transmission efficiency because the derived attention scores push the transmission resource allocated to semantic significant parts.

The redundancy removal module consists of a FC layer, which outputs a probability matrix of size ${B \times q \times 1001}$, and each element of the last dimension represents a token in the vocabulary list. The vocabulary list contains 1000 subwords and a special token, which represents both the symbol $EOS$, and $BOS$, where $EOS$ marks the end of a sentence, and $BOS$ is used at the beginning of a sentence. We note here we use the subwords as tokens instead of the words as in our previous work \cite{han2022semantic}, which reduces the size of the vocabulary from 15003 (the most frequent 15000 words and three special symbols) to 1001. With the output of the FL layer, which indicates the corresponding subword of each element in the input sequence $\boldsymbol Z$, the redundancy removal module removes from $\boldsymbol Z$ the special tokens and the sequences after the first special token in the sequence excepting the one that may be at the beginning of the sequence, which represents the symbols $EOS$. This is because the parts after $EOS$ are without any semantic meaning. The experiments reveal that cutting off the sequences after the symbol $EOS$ saves about 59.4\% of the transmission length, and removing the special saves approximately 4.5\%. We note that the output of the FC layer in this redundancy removal module is also fed into an embedding layer to get an embedding of size ${B \times q \times 128}$ to be concatenated with  $\boldsymbol H^{\prime}$ of the next time stamp before feeding into the LSTM layer in the soft alignment module, which passes the semantic information from the current time stamp to the next one.

\subsubsection{Channel Encoder and Channel Decoder}
In the channel encoder at the transmitter side, two cascaded FC layers map $\boldsymbol L$ to symbol sequences $\boldsymbol X \in\mathfrak R^{B \times c \times 64}$, which is then reshaped into $\boldsymbol X^{\prime} \in\mathfrak R^{B \times 128c \times 2}$, where the first and second channels are the real parts and imaginary parts of the wireless signals to be transmitted, respectively. The received symbol sequences, $\boldsymbol Y \in\mathfrak R^{B \times 32c \times 2}$ at the receiver, are first reshaped into $\boldsymbol Y^{\prime} \in\mathfrak R^{B \times c \times 64}$, which are then input to two cascaded FC layers to recover the text-related semantic features, denoted by $\widehat {\boldsymbol L} \in\mathfrak R^{B \times c \times 1024}$. $\widehat {\boldsymbol L}$ is then fed into the semantic decoder to recover the transcription of the input speech signals. We note here, in contrast to the conventional channel coding, the proposed channel encoder reduces the dimension of the transmit symbols instead of introducing redundancy to combat. This is because the proposed learning based channel encoder and decoder mitigate the effect of channel noises by adapting the neural network weights through end to end optimization. 

\subsubsection{Beam Search Decoder and Semantic Corrector}
To successfully predict the corresponding transcriptions, We introduce a beam search decoder with a semantic corrector that obtains more accurate results by exploiting both internal and external semantic knowledge. The beam search decoder operates as follows. It starts with the beginning of the input sequence, corresponding to the special token \textit{SOS}, and calculates the conditional distribution of every subword in the vocabulary, which contains 1001 tokens in our case, given \textit{SOS}. Then $k$ most likely subwords are selected according to this conditional distribution, and added to the beam that forms $k$ partial transcriptions. Then the following step is repeated until the whole input sequence is processed: at each iteration it calculates $k$ conditional distributions over the vocabulary given each of the current $k$ partial transcriptions in the beam. According to these conditional distributions, we select the $k$ most likely partial transcriptions from all the possible extended transcriptions, that is $k\times 1001$ candidates, and replace the previous $k$ most likely partial transcriptions. At last, the beam search decoder outputs the most likely transcription in its beam. We note that the beam encoder exploits the long time dependency in the semantic sequence by keeping $k$ candidates in the beam and evaluating the conditional distribution given the previous context at each iteration, which improves the accuracy of the predicted transcriptions as revealed by the numerical results.

We also include a semantic corrector, which is based on a pretrained RNN-based language model, to further correct possible semantic errors by leveraging external semantic knowledge. In particular, this language model computes the corresponding distributions at each iteration of the beam search, which is combined with the distribution mentioned above with a specific weight. This combined distribution is then used to select the best $k$ partial transcription at each iteration. We note that the adopted language model has been trained on a large and comprehensive text dataset, hence containing substantial semantic information that helps prediction of the transcription. As the numerical results reveal, the semantic corrector improves the system's overall performance and corrects the semantic mistakes that may be introduced by the noisy channel.

\subsection{Speech to Speech Transmission}
For the speech to speech transmission, we further include an additional speech information extractor at the transmitter side and speech reconstructor at the receiver side on top of the subsystem for the speech to text transmission, as shown in Fig. \ref{dlmodel}, where the additional speech information extractor generates the additional speech related information, such as speech duration, pitch and power, to send over the channel, for better reconstruction of the speech signals, and the speech reconstructor synthesizes speech from the received text and additional speech information. We introduce each module in detail in the following.

\subsubsection{Additional Speech Information Extractor}
The additional speech information extractor takes the intermediate features $\boldsymbol H$ from the FC module in the semantic encoder as input and outputs the frame-level alignment to obtain duration information of the speech sequence ${\boldsymbol C \in\mathfrak R^{B \times c^{\prime} \times 1}}$, where $c^{\prime}$ is the total number of predicted phonemes, and each element in $\boldsymbol C$ represents the duration of each phoneme. In particular, a FC layer is employed to compute the probabilities of subwords for each frame in the sequence input. Then the CTC alignment module, based on the Viterbi algorithm \cite{forney1973viterbi}, decides the most possible subword for each frame, and outputs the start time and the ending time as the duration of each predicted subword. We also utilize a pretained subword to phoneme model \cite{g2pE2019} to acquire the corresponding phoneme of each subword. We also compute the pitch and power information of each phoneme from the input speech spectrums, which are concatenated with the duration information $\boldsymbol C$ to form the set of speech related information ${\boldsymbol D \in\mathfrak R^{B \times c^{\prime} \times 3}}$ to be sent to the receiver for the purpose of speech recovery. We note that compared to the semantic information to be sent, which is of dimension ${\boldsymbol L \in\mathfrak R^{B \times c \times 1024}}$, the amount of the additional speech information is neglectable, which is thus transmitted over the channel using the traditional channel coding scheme, and can be recovered losslessly at the receiver.

\subsubsection{Speech Reconstructor}
\begin{figure}[tbp]
\includegraphics[width=0.9\textwidth]{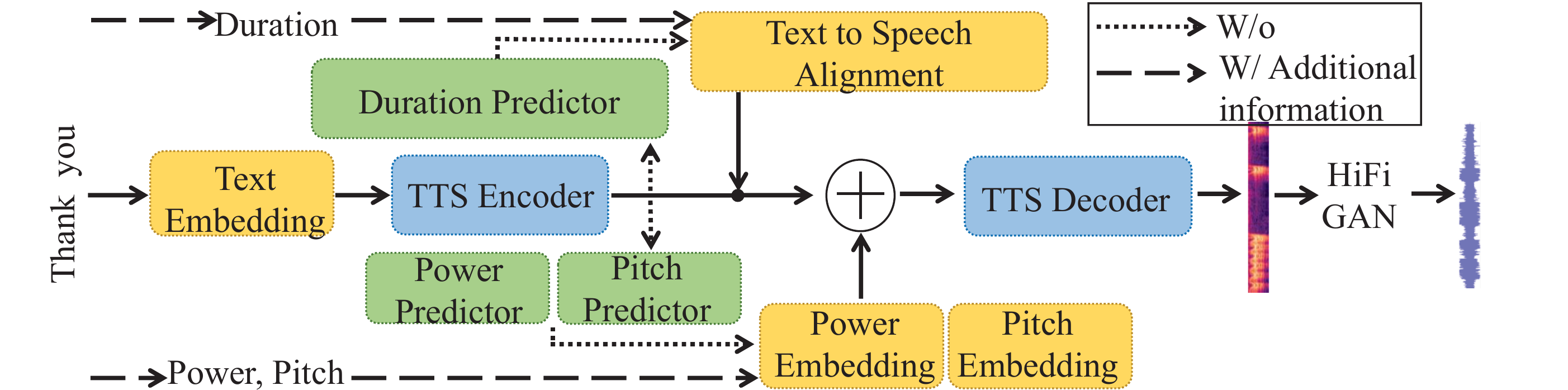} % image name
\centering % central position
\caption{Adopted FastSpeech2 Speech Reconstructor}  % figure title
\label{speechtospeech}  
\end{figure}

The speech reconstructor, as shown in Fig.\ref{speechtospeech}, takes the predicted transcription from the semantic decoder and corrector as input and outputs the recovered speech signal with the help of the additional speech information from the transmitter. In particular, a text embedding layer is first employed to transform the input text to an embedding, which is then fed to a text-to-speech(TTS) encoder containing four transformer layers to obtain a latent representation of each token in the input text sequence, together with the prediction of the duration, power and pitch of the corresponding phonemes. Then a text to speech alignment module is exploited that takes in the predicted duration information and the real duration information received from the receiver, and modifies the number of frames of each token in the latent representation output by the TTS encoder. This latent representation is then combined with the power information embedding and pitch information embedding, which embeds both the transmitted and predicted power and pitch information, respectively, as shown in Fig.\ref{speechtospeech}, to generate latent representation that contains both the semantic relevant text information and speech-related information. This combined information is then input to the TTS decoder, which consists of 6 transformer layers, to recover the speech spectrums. Finally, we use a pretrained GAN, that is, HiFiGAN vocoder \cite{kong2020hifi}, to generate the speech signals from the recovered spectrums.

\subsection{Model Training and Testing} 

In this section, we describe the training and testing of the proposed DL-based semantic communication system. 

\subsubsection{Data Argument Strategies}
We employ data argument strategies to enrich the training dataset in order to improve the performance of the trained model. In particular, we exploit the following two data argument strategies: we speed up the original speech to 1.1$\times $ and slow down the original speech to 0.95$\times$ to obtain two times more training data \cite{ ko2015audio}; we simulate environment noises randomly with SNR of 0dB to 15dB using room impulse responses and add these noises to the input speech signals to generate new speech data\cite{ko2017study}.

\subsubsection{Two-stage Training} 
We use a two-stage training method, which converges faster than training the network as a whole in an end-to-end manner, as revealed by experiments. In the first stage, we train the semantic encoder, the semantic decoder, and the fully connected layer in the additional speech information extractor while ignoring all the other parts including the noisy channel by directly inputting the output of the semantic encoder to the semantic decoder. We train these parts jointly with the following loss function using the Adadelta optimizer:
\begin{equation} {\mathcal L}_{all} = \lambda {\mathcal{L}_{CTC}+(1-\lambda){\mathcal{L}}_{ce}}, \label{all} \end{equation}
where $\mathcal{L}_{CTC}$ denote the CTC loss between the true token distribution in the input speech and the output of the FC layer in the additional information extractor, which is given by:
\begin{equation} {\mathcal{L}_{CTC}=-\ln \left(\prod {p}\left({\boldsymbol J} \mid \boldsymbol{S} \right)\right) }, \label{CTC}  \end{equation}
which calculates the product of the probability distribution of output alignment $\boldsymbol J$ by the FC layer given the input spectrum $\boldsymbol S$; and ${\mathcal{L}}_{ce}$ denotes the  cross-entropy loss between the predicted transcription $\widehat {\boldsymbol G}$, and the ground truth for the transcription $\boldsymbol G$, which is given by
\begin{equation} {\mathcal L}_{ce}(\boldsymbol G, \widehat{\boldsymbol G})=- \sum^{m}_{i=1} G_{i}\log(\widehat{{G}_{i}}). \label{cross_entropy}  \end{equation}
Due to the presence of a feedback loop in the soft alignment module, we use the teacher forcing strategy \cite{williams1989learning} to achieve fast convergence. Specifically, at the beginning of the training, as the output of the FC layer is incorrect, we use the true transcriptions instead to speed up the learning of correct alignment.

In the second stage, we freeze the trained semantic encoder, and the FC layer in the additional information extractor, and train other parts of the network, including the channel encoder, channel decoder, and semantic decoder under physical channel with random SNR. The cross-entropy loss function and the Adadelta optimizer are also utilized. For \textit{speech to speech transmission} task, we further minimize the mean square error (MSE) loss between the input spectrum $\boldsymbol S$ and the recovered spectrum $\widehat{\boldsymbol S}$, given by
\begin{equation} {\mathcal L}_{mse}(\boldsymbol S, \widehat{\boldsymbol S})= \sum (\boldsymbol S-\widehat{\boldsymbol S})^2. \label{MSE}  \end{equation}
We note here a pretrained model, HiFiGAN\cite{ren2020fastspeech, kong2020hifi}, is employed for the speech reconstructor, which does not participate in the training. The first and second stages of training are described in Algorithm \ref{Training algorithm1} and \ref{Training algorithm2}, respectively.

\subsubsection{Testing Stage}
The testing algorithms for the \textit{speech to text} transmission and the \textit{speech to speech} transmission are illustrated in Algorithm \ref{testing algorithm} and \ref{speech testing algorithm}, where we note that the speech sample sequences used for testing are different from that used for training.

\begin{algorithm}[htb]
\caption{Training algorithm of the first stage.}
\label{Training algorithm1}

\begin{algorithmic}[1]  
    \State  \textbf{Input:}  Speech signals and transcriptions $\boldsymbol G$ from dataset %fading channel $\boldsymbol h$, noise $\boldsymbol w$.
    
        \While{Stop criterion is not meet}
            \State Generate spectrum sequences $\boldsymbol S $ from input speech signals.
            \State Output $\boldsymbol L$ and $\boldsymbol H$ from $\boldsymbol S $ by the semantic encoder.
            \State Output $\widehat {\boldsymbol G}$ from $\boldsymbol L$ by the semantic decoder
            \State Compute $l_{ce}$ via (\ref{cross_entropy})
            \State Output ${\boldsymbol J}$ from $\boldsymbol H$ by the FC layer of the additional speech information extractor.
            \State Compute $l_{ctc}$ via (\ref{CTC})
            \State Compute $l_{all}$ from $l_{ce}$ and $l_{ctc}$ via (\ref{all})
            \State Update the weight of the model via Adadelta.
        \EndWhile
	\State \textbf{Output:} Trained semantic encoder and decoder.
\end{algorithmic}

\end{algorithm}

\begin{algorithm}[htb]
\caption{Training algorithm of the second stage.}
\label{Training algorithm2}

\begin{algorithmic}[1]  
    \State  \textbf{Input:}  Speech signals and transcriptions $\boldsymbol G$ from dataset, fading channel $\boldsymbol h$, noise $\boldsymbol w$, trained semantic encoder, trained semantic decoder. %and pretained speech reconstructor.
    
    \State Set fading channel $\boldsymbol h$ = Rayleigh or AWGN
    \State freeze the trained semantic encoder
        \While{Stop criterion is not meet}
            \State Generate Gaussian noise $\boldsymbol W$ under random SNR value for every epoch.
            \State Generate spectrum sequences $\boldsymbol S $ from input speech signals.
            \State Output $\boldsymbol L$ from $\boldsymbol S $ by the semantic encoder.
            \State Output $\boldsymbol X$ from $\boldsymbol L^{\prime}$ by the channel encoder.
            \State Transmit $\boldsymbol X$ and receive $\boldsymbol Y$ via (\ref{channel}).
            \State Output $\widehat{\boldsymbol L}$ from $\boldsymbol Y$ by the channel decoder.
            \State Output $\widehat {\boldsymbol G}$  from $\widehat{\boldsymbol L^{\prime}}$ by the semantic decoder.
            %\State Output the recovered speech signal form $\widehat {\boldsymbol G}$ with the speech reconstrutor. 
            \State compute $l_{ce}$ via (\ref{cross_entropy}) % and mse loss
            \State Update the weight of the model via Adadelta.
        \EndWhile
    \State Finetuning the pretained speech reconstructor withthe received text, the addition speech information and the orignal speech signals vie ${\mathcal L}_{mse}$ (\ref{MSE}).
	\State \textbf{Output:} Trained semantic decoder, channel encoder, channel decoder, and speech reconstructor
\end{algorithmic}

\end{algorithm}

%%%%%%%%%%     Algorithm: testing algorithm     %%%%%%%%%%%%
\begin{algorithm}[htb]
\caption{Testing algorithm of the proposed speech to text transmission.}
\label{testing algorithm}

\begin{algorithmic}[1]  
    \State  \textbf{Input:}  Speech signals and transcriptions $\boldsymbol G$ from dataset, fading channel $\boldsymbol h$, noise $\boldsymbol w$.
    
    \State Set fading channel $\boldsymbol h$ = Rayleigh or AWGN
        \For{each SNR value}
            \State Generate Gaussian noise $\boldsymbol W$ under the SNR value.
            \State Generate spectrum sequences $\boldsymbol S $ from input speech signals.
            \State Output $\boldsymbol L$ from $\boldsymbol S $ by the semantic encoder.
            \State Output $\boldsymbol X$ from $\boldsymbol L^{\prime}$ by the channel encoder.
            \State Transmit $\boldsymbol X$ and receive $\boldsymbol Y$ via (\ref{channel}).
            \State Output $\widehat{\boldsymbol L}$ from $\boldsymbol Y$ by the channel decoder.
            \State Output $\widehat {\boldsymbol G}$  from $\widehat{\boldsymbol L^{\prime}}$ by the semantic decoder.
        \EndFor
	\State \textbf{Output:} compare $\widehat{\boldsymbol G}$ and $\boldsymbol G$ and compute WER scores and sentence similarity via (\ref{WER}) and (\ref{similar})
\end{algorithmic}

\end{algorithm}

\begin{algorithm}[htb]
\caption{Testing algorithm of the proposed speech to speech transmission.}
\label{speech testing algorithm}

\begin{algorithmic}[1]  
    \State  \textbf{Input:}  Speech signals from dataset, fading channel $\boldsymbol h$, noise $\boldsymbol w$.
    
    \State Set fading channel $\boldsymbol h$ = Rayleigh or AWGN
        \For{each SNR value}
            \State Generate Gaussian noise $\boldsymbol W$ under the SNR value.
            \State Generate spectrum sequences $\boldsymbol S $ from input speech signals.
            \State Output $\boldsymbol L$ and $\boldsymbol C$ from $\boldsymbol S $ by the semantic encoder.
            \State Output $\boldsymbol X$ from $\boldsymbol L$ by the channel encoder.
            \State Output $\boldsymbol D$ from $\boldsymbol C$ by the Additional speech information extractor
            \State Transmit $\boldsymbol X$ and receive $\boldsymbol Y$ via (\ref{channel}).
            \State Transmit $\boldsymbol D$ in a lossless manner. 
            \State Output $\widehat{\boldsymbol L}$ from $\boldsymbol Y$ by the channel decoder.
            \State Output $\widehat {\boldsymbol G}$  from $\widehat{\boldsymbol L^{\prime}}$ by the semantic decoder.
            \State Output the recovered speech signal form $\widehat {\boldsymbol G}$ and $\boldsymbol D$ with the speech reconstructor.
        \EndFor
	\State \textbf{Output:} compare the recovered speech signal and input speech signal and compute MCD with DTW via (\ref{PESQ}).  Mos can be tested using Amazon Mturk.
\end{algorithmic}
\end{algorithm}

\section{Experiment and Numerical Results}

In this section, we compare the proposed method's performance to the existing deep learning-based semantic communication systems for speech transmission. We consider the AWGN and Rayleigh channels for the evaluation. We use the Librispeech dataset\cite{panayotov2015librispeech} for training and testing, which is a speech to text library based on public domain audiobooks containing 960 hours of speech for training and 2703 utterances for testing. For the speech to text transmission, we compare to the following three benchmarks: the existing DL-based speech to text transmission approach by \cite{weng2021semantic}, referred to as DeepSC-SR; the semantic communication approach for the text to text transmission proposed in \cite{qin1}, which transmits and recovers text at the receiver, combined with the proposed semantic encoder, which transfers the speech signal to semantic text, referred to as SE-DeepSC; the semantic communication for speech transmission proposed in \cite{Weng2101:Semantic}, which transmits and recovers speech, combined with the proposed semantic encoder, which transfers the received speech signal to text at the receiver, referred to as DeepSC-S-SR. For a fair comparison, we use the same dimension of the input spectrum and training data augmentation strategies. We also use the proposed semantic encoder and decoder, while ignoring the noisy channel as well as channel encoder and decoder, as the Baseline, which provides the upper-bound performance. For the speech to speech transmission, we also compare with the speech transmission scheme proposed in \cite{Weng2101:Semantic}, but without the proposed semantic encoder applied, referred to as DeepSC-S. The detailed setting of our proposed network is shown in table \ref{table 2}, which lists the neural layers and the numbers of parameters in each module, as well as the type of activation functions applied. We note that the proposed approaches and all the methods for comparison are trained and tested with the same datasets. During training, we set channel conditions with SNR varying randomly from 5dB to 10dB.

\begin{table}
\centering
\footnotesize
\caption{Parameter settings of the proposed semantic communication Network for speech recognition}
\label{table 2}
\begin{tabular}{|c|ccc|c|}
\hline
                                                                                            & \multicolumn{2}{c|}{Layer Name}                                                                                                                                                                            & Parameters                                      & Activation            \\ \hline
\multirow{18}{*}{Semantic Encoder} & \multicolumn{1}{c|}{\multirow{5}{*}{VGG Module}}                                           & \multicolumn{1}{c|}{2$\times$CNN}                    & 3$\times$3/128                                  & LeakyReLU             \\ \cline{3-5} 
                                                                                            & \multicolumn{1}{c|}{}                                                                                                                               & \multicolumn{1}{c|}{MaxPool}                         & 2$\times$2                                      & None                  \\ \cline{3-5} 
                                                                                            & \multicolumn{1}{c|}{}                                                                                                                               & \multicolumn{1}{c|}{2$\times$CNN}                    & 3$\times$3/256                                  & LeakyReLU             \\ \cline{3-5} 
                                                                                            & \multicolumn{1}{c|}{}                                                                                                                               & \multicolumn{1}{c|}{MaxPool}                         & 2$\times$2                                      & None                  \\ \cline{3-5} 
                                                                                            & \multicolumn{1}{c|}{}                                                                                                                               & \multicolumn{1}{c|}{Reshape}                         & None                                            & None                  \\ \cline{2-5} 
                                                                                            & \multicolumn{1}{c|}{\multirow{2}{*}{BLSTM Module}}                                         & \multicolumn{1}{c|}{\multirow{2}{*}{4$\times$BLSTM}} & \multirow{2}{*}{1024} & \multirow{2}{*}{None} \\
                                                                                            & \multicolumn{1}{c|}{}                                                                                                                               & \multicolumn{1}{c|}{}                                &                                                 &                       \\ \cline{2-5}
                                                                                            & \multicolumn{1}{c|}{\multirow{2}{*}{FC}}                                                                                                            & \multicolumn{1}{c|}{FC}                              & 1024                                            & LeakyRelu             \\ \cline{3-5} 
                                                                                            & \multicolumn{1}{c|}{}                                                                                                                               & \multicolumn{1}{c|}{FC}                              & 1024                                            & LeakyRelu             \\ \cline{2-5} 
                                                                                            
                                                                                            & \multicolumn{1}{c|}{\multirow{7}{*}{ Soft Alignment Module}} & \multicolumn{1}{c|}{Query FC}                        & 300                                             & None                  \\ \cline{3-5} 
                                                                                            & \multicolumn{1}{c|}{}                                                                                                                               & \multicolumn{1}{c|}{Key FC}                          & 300                                             & None                  \\ \cline{3-5} 
                                                                                            & \multicolumn{1}{c|}{}                                                                                                                               & \multicolumn{1}{c|}{Conv1d}                          & 201/10/100                                      & None                  \\ \cline{3-5} 
                                                                                            & \multicolumn{1}{c|}{}                                                                                                                               & \multicolumn{1}{c|}{FC}                              & 300                                             & None                  \\ \cline{3-5} 
                                                                                            & \multicolumn{1}{c|}{}                                                                                                                               & \multicolumn{1}{c|}{FC}                              & 1                                               & Tanh                  \\ \cline{3-5} 
                                                                                            & \multicolumn{1}{c|}{}                                                                                                                               & \multicolumn{1}{c|}{GRU}                             & 1024                                            & None                  \\ \cline{3-5} 
                                                                                            & \multicolumn{1}{c|}{}                                                                                                                               & \multicolumn{1}{c|}{Embedding}                       & 1000/128                                        & None                  \\ \cline{2-5} 
                                                                                            & \multicolumn{1}{c|}{ Redandancy Removal Module}              & \multicolumn{1}{c|}{FC}                              & 1001                                            & logsoftmax            \\ \cline{2-5} 
                                                           &                                  \multicolumn{1}{c|}{CTC alignment Module}                                                                                                           & \multicolumn{1}{c|}{FC}                              & 1003                                            & logsoftmax            \\ \hline
\multirow{3}{*}{Channel Encoder}   & \multicolumn{1}{c|}{\multirow{2}{*}{FC}}                                                                                                            & \multicolumn{2}{c|}{256}                                                             & ReLU                  \\ \cline{3-5} 
                                                                                            & \multicolumn{1}{c|}{}                                                                                                                               & \multicolumn{2}{c|}{64}                                                              & None                  \\ \cline{2-5} 
                                                                                            & \multicolumn{2}{c|}{Reshape}                                                                                                                                                                               & None                                            & None                  \\ \hline
\multirow{3}{*}{Channel Decoder}   & \multicolumn{2}{c|}{Reshape}                                                                                                                                                                               & None                                            & None                  \\ \cline{2-5} 
                                                                                            & \multicolumn{1}{c|}{\multirow{2}{*}{FC}}                                                                                                            & \multicolumn{2}{c|}{64}                                                              & LeakyRelu             \\ \cline{3-5} 
                                                                                            & \multicolumn{1}{c|}{}                                                                                                                               & \multicolumn{2}{c|}{256}                                                             & LeakyRelu             \\ \hline
Semantic Decoder                   & \multicolumn{2}{c|}{FC}                                                                                                                                                                                    & 1000                                            & LogSoftmax            \\ \hline
\multirow{3}{*}{semantic corrector}                                                         & \multicolumn{2}{c|}{embedding}                                                                                                                                                                             & 128                                             & None                  \\ \cline{2-5} 
                                                                                            & \multicolumn{2}{c|}{2$\times$LSTM}                                                                                                                                                                                & 2048                                            & None                  \\ \cline{2-5} 
                                                                                            & \multicolumn{2}{c|}{FC}                                                                                                                                                                                    & 512                                             & LeakyRelu             \\ \hline
\multirow{7}{*}{speech reconstrutor}                                                        & \multicolumn{2}{c|}{4$\times$tranformer}                                                                                                                                                                          & 256(2 heads)                                    & None                  \\ \cline{2-5} 
                                                                                            & \multicolumn{2}{c|}{text embedding}                                                                                                                                                                   & 256                                             & None                  \\ \cline{2-5} 
                                                                                            & \multicolumn{2}{c|}{pitch embedding}                                                                                                                                                                       & 256                                             & None                  \\ \cline{2-5} 
                                                                                            & \multicolumn{2}{c|}{power embedding}                                                                                                                                                                       & 256                                             & None                  \\ \cline{2-5} 
                                                                                            & \multicolumn{2}{c|}{speaker embedding}                                                                                                                                                                     & 256                                             & None                  \\ \cline{2-5} 
                                                                                            & \multicolumn{2}{c|}{6$\times$transformer}                                                                                                                                                                          & 256(2 heads)                                    & None                  \\ 
                                                                                            \hline
\end{tabular}
\end{table}

\subsection{Performance comparison of the speech to text transmission}

We compare the text prediction performance of the proposed approaches with the aforementioned three benchmarks, the baseline with and without the proposed semantic corrector module, denoted by \textit{Baseline w/ semcor} and \textit{Baseline w/o semcor}, respectively, and our previous approach \cite{han2022semantic} as well. We present the performance comparison in terms of WER and sentence similarity in Fig. 6 and Fig. 7, respectively. We can see that the proposed method significantly outperforms the other methods under both channel conditions in terms of both WER and sentence similarity. We can also note that our proposed method performs steadily and closer to the baseline, while the performance of the SE-DeepSC approach is poor under a low SNR regime. We note that we get the WER of around 15\% instead of 40\% for DeepSC-SR as reported in the original paper \cite{weng2021semantic}, which may benefit from the use of 40 dimension fbank features and the data argument strategies. It can be observed from both figures that SE-DeepSC performs not as well as the other two counterparts. This may be due to the semantic errors in the output of the semantic encoder, which are neglected in the transmission and reconstruction, and hence, exist in the final output. It can also be observed that DeepSC-S-SE and DeepSC-SR also show a certain level of robustness to the channel noises, which may benefit from the end to end training.

We can observe a significant improvement over our previous approach, which may come from the following several aspects. First is the use of subwords as tokens instead of the natural words, which avoids the errors of unseen words. Then it may also be due to the data argument strategies employed, which enriches the training dataset by several times. The other reason is the introduction of the beam search decoder and the semantic corrector, which fixes semantic errors utilizing the internal and external semantic knowledge. The benefit of semantic corrector is validated by the results shown in Fig. 6 and Fig. 7, where we can see a clear performance gap between the baseline with and without semantic corrector. And the usage of greedy semantic decoder, denoted as Greedy semdec, shows worse performance than both our proposed methods.
We further present text output examples by greedy decoder used by DeepSC-SR and our semantic decoder with and without the semantic corrector in Table. \ref{tablelongforabalation}, where highlight the mistaken parts with different colors for different methods. It can be seen that the beam search semantic decoder avoids some word errors occurring to the greedy decoder, while the semantic corrector further fixes the other mistakes, and pushes the output closer to ground truth. We further present the performance comparison between the proposed approach with and without semantic corrector, and also the proposed approach with the semantic decoder replaced with the greedy decoder, in terms of WER and sentence similarity over different channels in Fig. \ref{abolationresult}. The results imply that the beam search decoder performs closely to the greedy semantic decoder, while the semantic corrector brings remarkable performance gain under different channels in terms of both the WER and sentence similarity.

\begin{figure}[tbp]
\centering % central position
  % figure title
 %\label{WER}
\subfigure[AWGN channel]{
\begin{minipage}{0.48\textwidth}
    \centering
    \includegraphics[width=\textwidth]{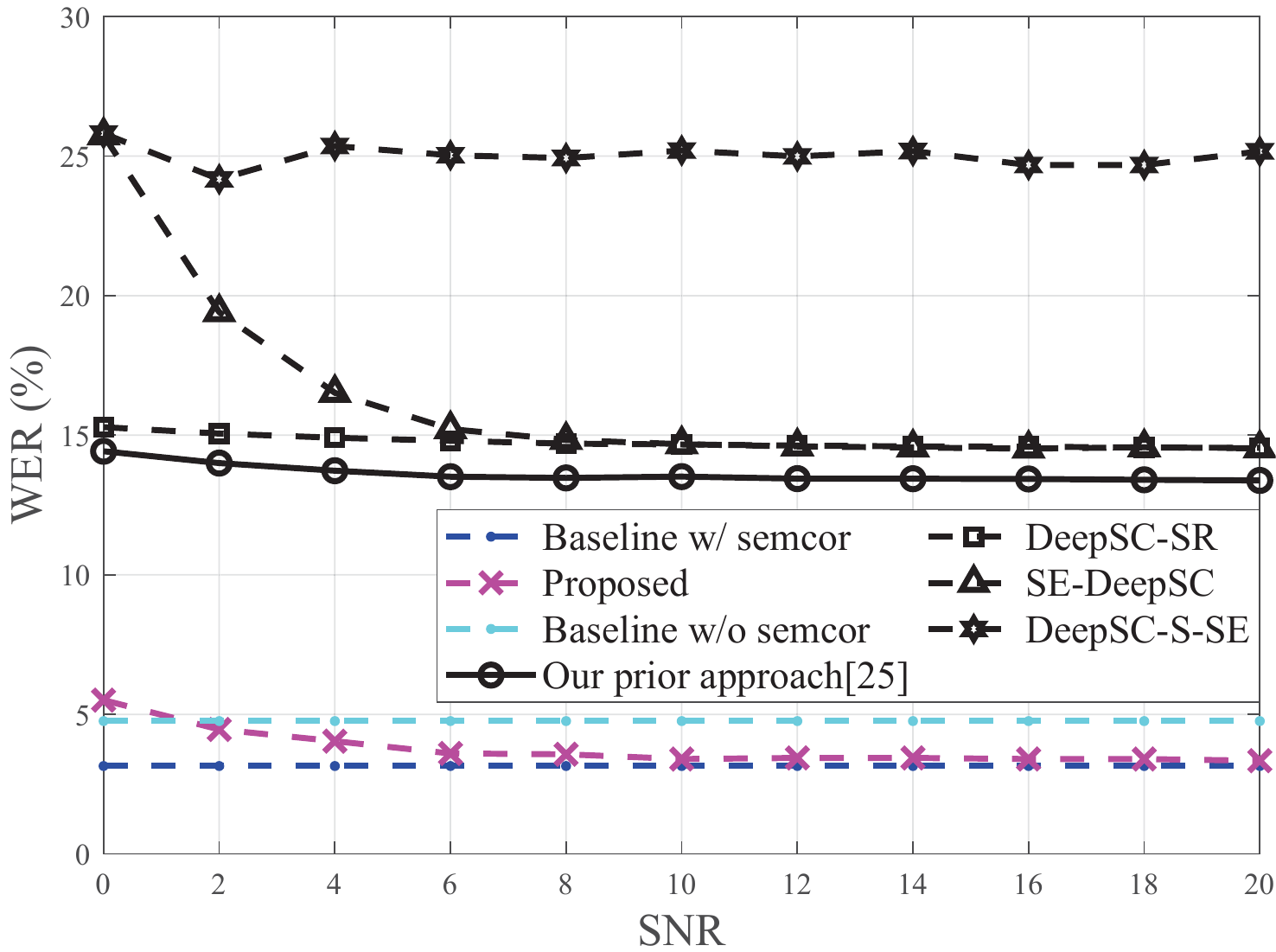}
    %\subcaption{AWGN channel}
    \label{WER_result_awgn}
\end{minipage}}
%\qquard
\subfigure[Fading channel]{
\begin{minipage}{0.48\textwidth}
    \centering
    \includegraphics[width=\textwidth]{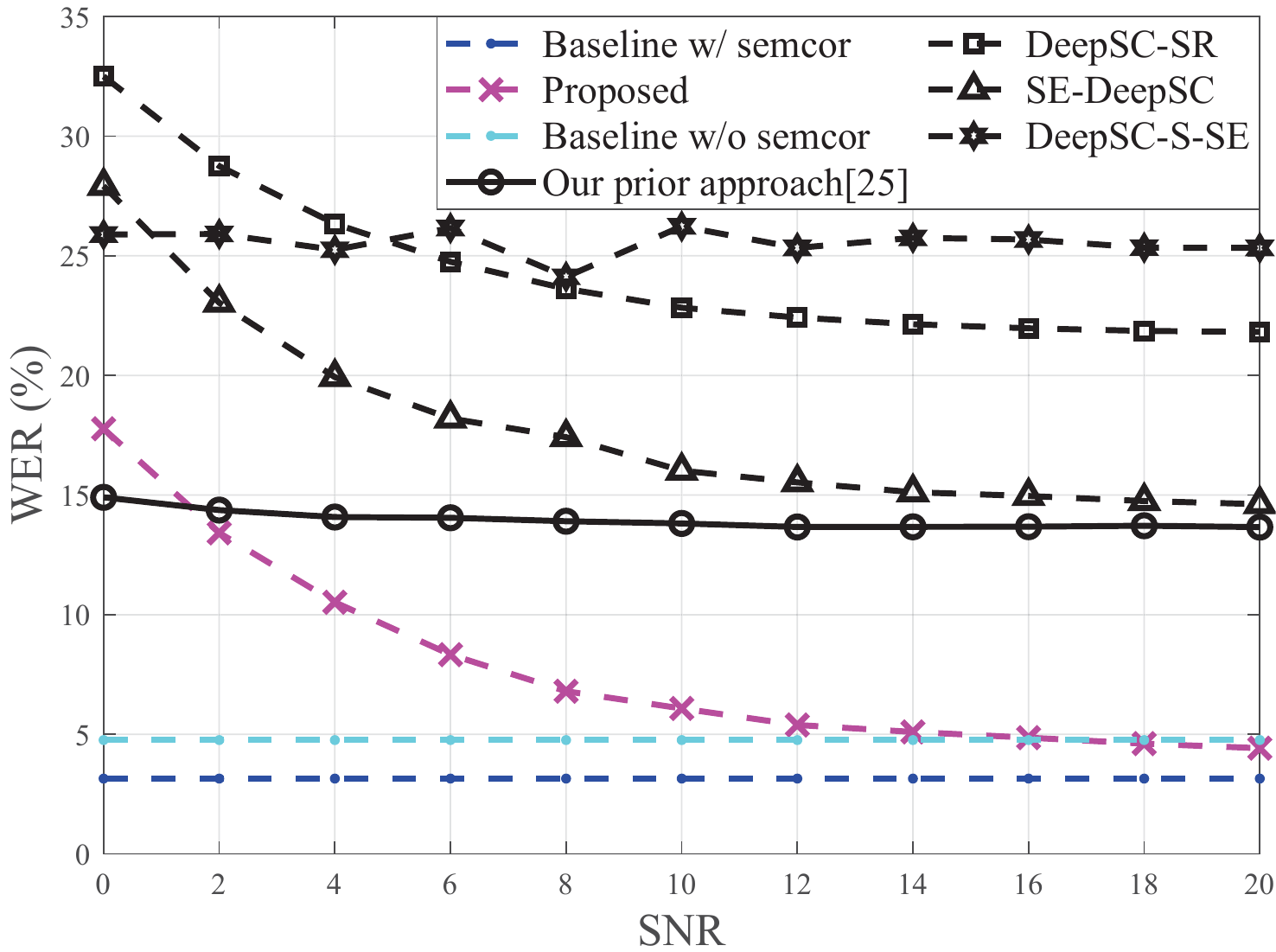}
    %\subcaption{Fading channel }
    \label{WER_result_fading}
\end{minipage}}
\caption{WER score versus SNR under different channels for different approaches. }
\end{figure}

\begin{figure*}
\centering % central position
  % figure title
  \subfigure[AWGN channel]{
\begin{minipage}{0.48\textwidth}
    \centering
    \includegraphics[width=\textwidth]{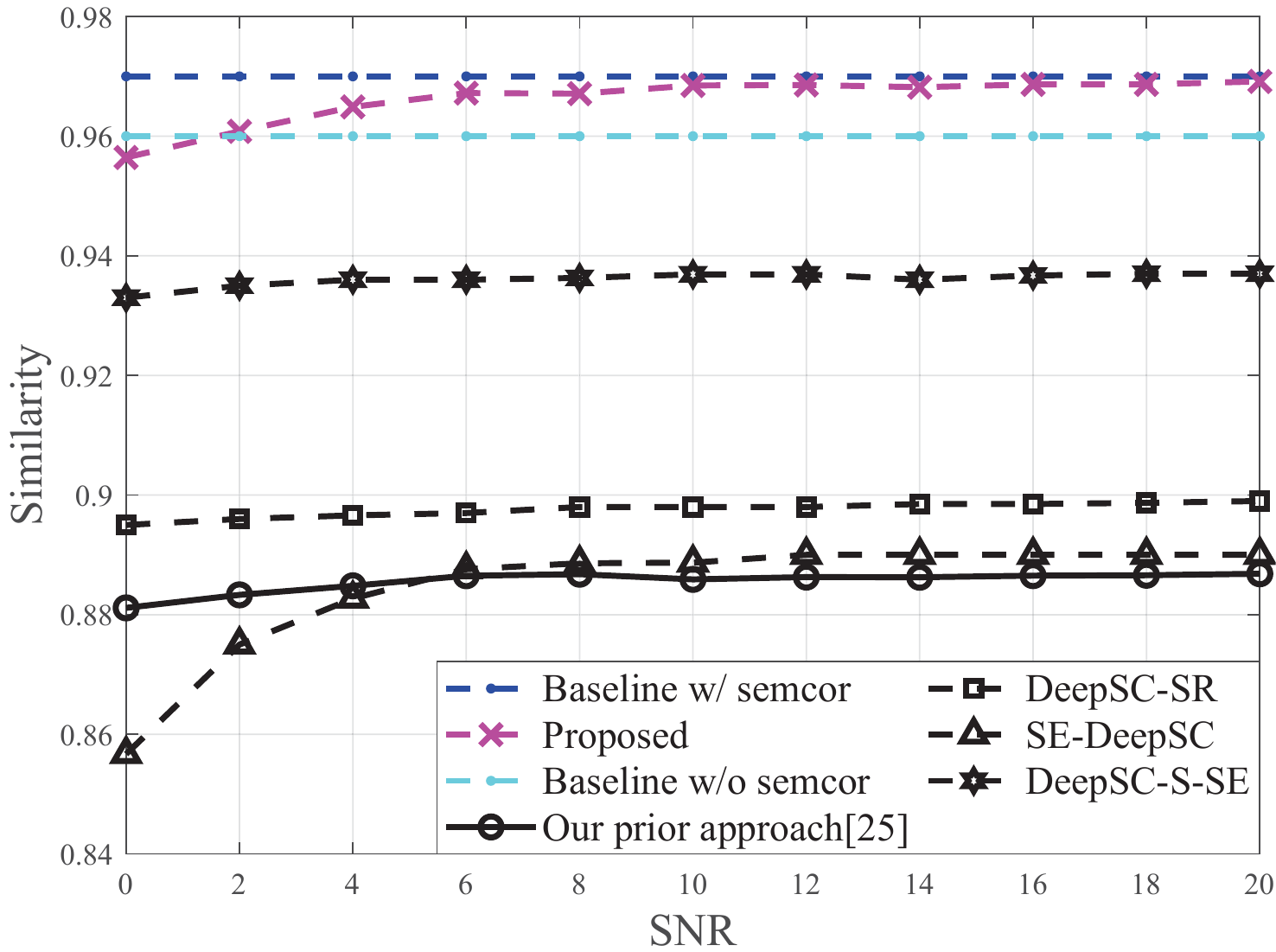}
    %\subcaption{AWGN channels}
    \label{sim_result_awgn}
\end{minipage}}
%\qquard
\subfigure[Fading channel]{
\begin{minipage}{0.48\textwidth}
    \centering
    \includegraphics[width=\textwidth]{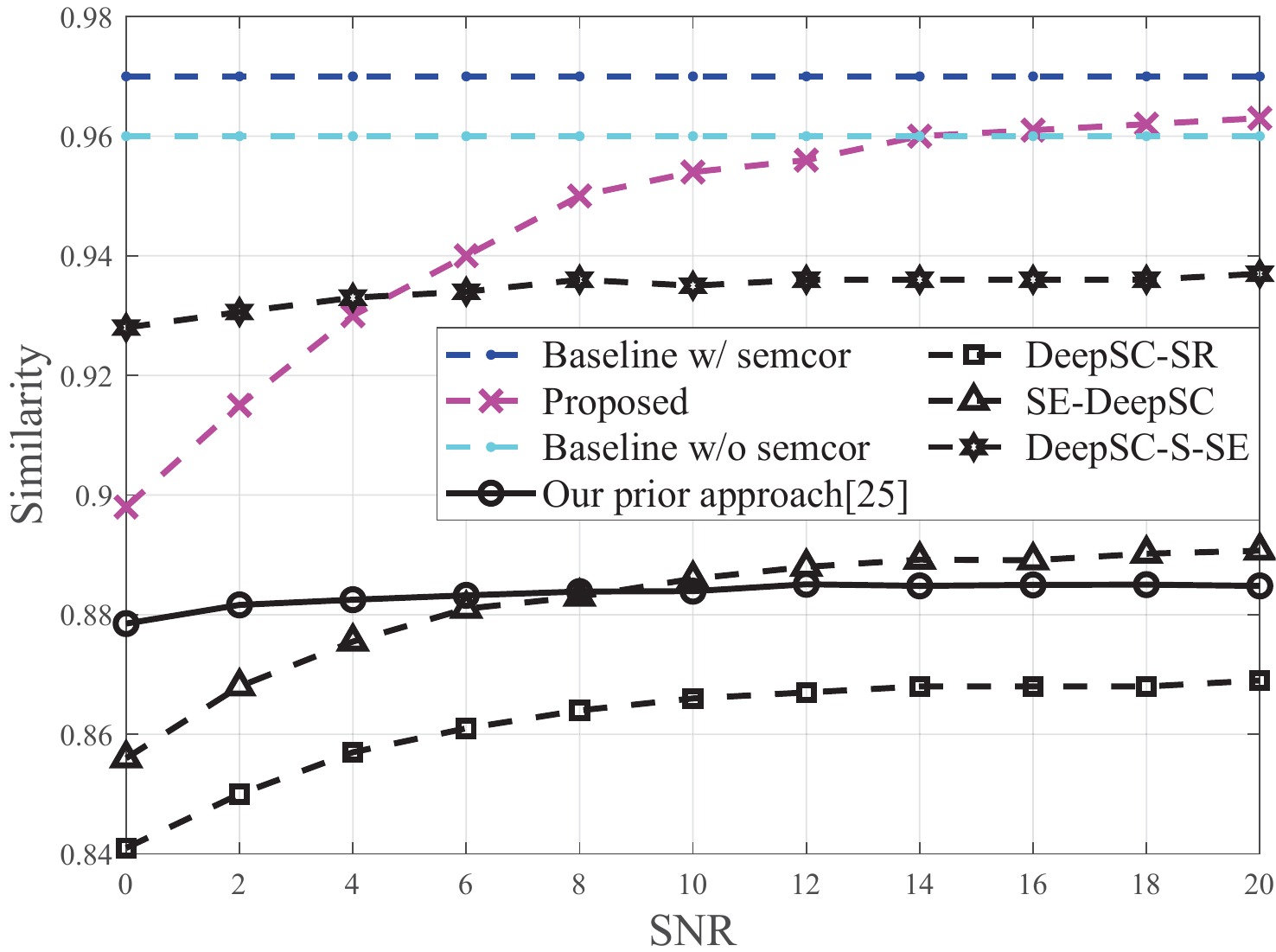}
    %\subcaption{Fading channels}
    \label{sim_result_fading}
\end{minipage}}
\caption{Sentence Similarity versus SNR Under different channels for different approaches.}
\end{figure*}

\begin{figure}
\centering % central position
  % figure title
\subfigure{
\begin{minipage}{0.48\textwidth}
    \centering
    \includegraphics[width=\textwidth]{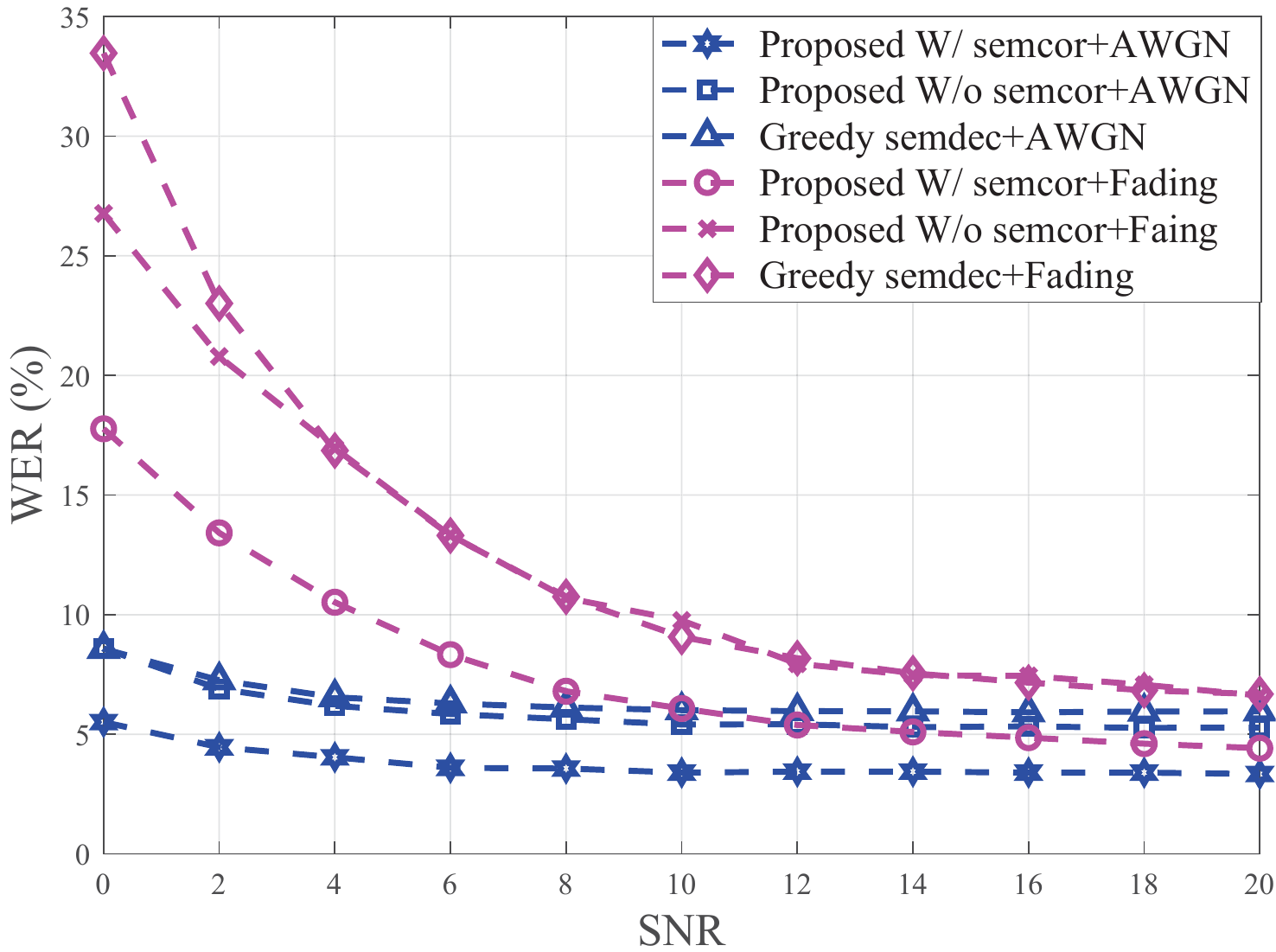}
    \label{werabalation}
\end{minipage}}
%\qquard
\subfigure{
\begin{minipage}{0.48\textwidth}
    \centering
    \includegraphics[width=\textwidth]{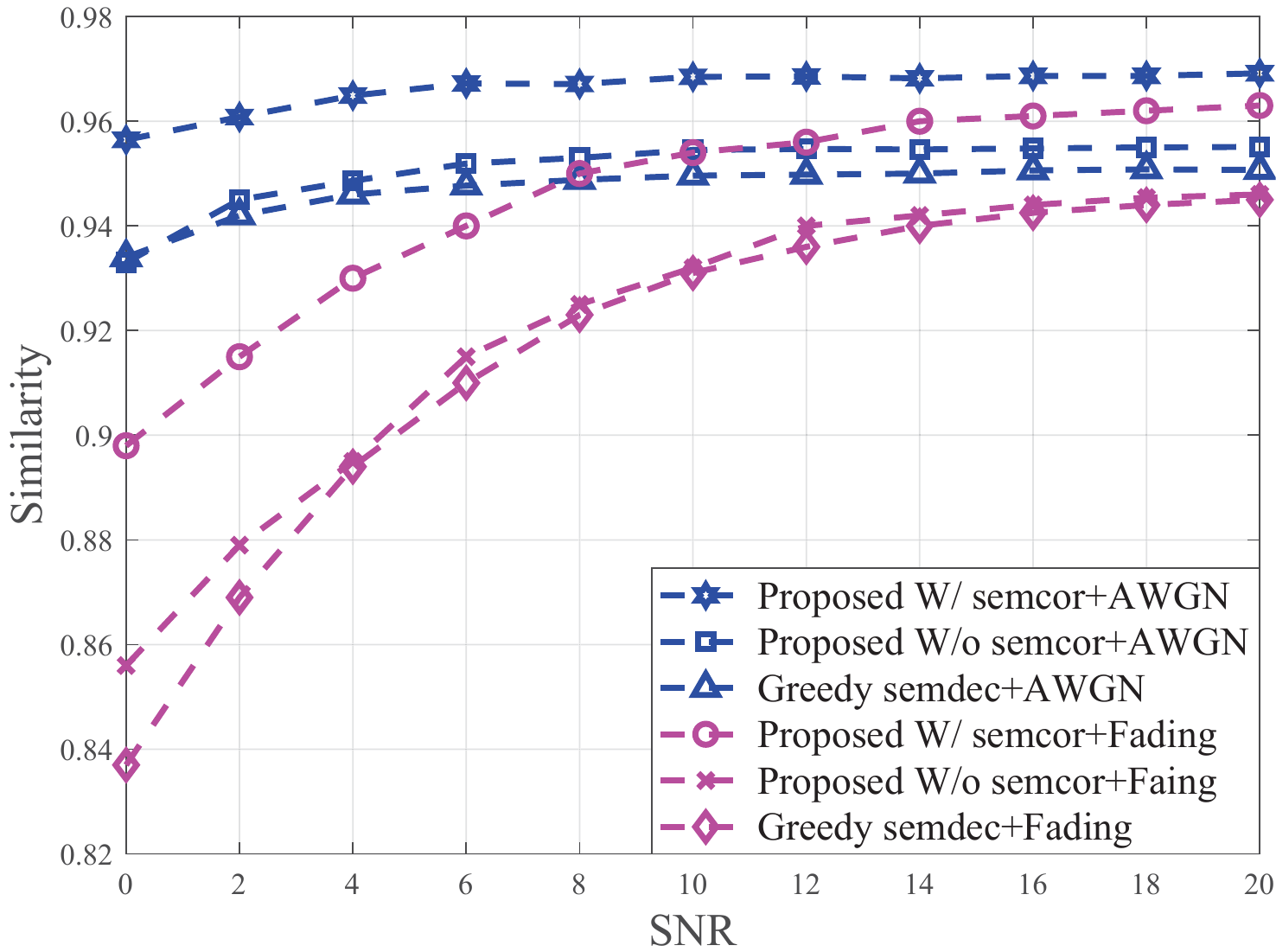}
    \label{simabalation}
\end{minipage}}
\caption{WER Score and Sentence Similarity versus SNR under different channels}  
\label{abolationresult}% figure title
\end{figure}

\begin{table}
\centering
  \caption{Examples of recovered transcriptions by different approaches.}
  \label{tablelongforabalation}
\begin{tabular}{|l|l|}
\hline
\diagbox{Method}{Ground Truth}                                    & MAYBE THEY'RE UP TO SOME OF THEIR GAMES AND WANTS ROUSING                                                                                                                                                                            \\ \hline
Greedy decoder                                               & \sethlcolor{yellow}\hl{MAY BE} \sethlcolor{yellow}\hl{THERE} UP TO SOME OF THEIR GAMES AND \sethlcolor{yellow}\hl{ONS} ROUSING \\ \hline
Beam search decoder                                          & \sethlcolor{pink}\hl{MAY BE} \sethlcolor{pink}\hl{THERE} UP TO SOME OF THEIR GAMES AND WANTS ROUSING                                                             \\ \hline
\multirow{2}{10em}{Beam search decoder with semantic corrector} & \multirow{2}{*}{MAYBE THEY'RE UP TO SOME OF THEIR GAMES AND WANTS ROUSING}                                                                                                                                                           \\
                                                             &                                                                                                                                                                                                                                      \\ \hline
\end{tabular}
\end{table}

\begin{table}
\centering
\caption{}
\label{table_number}
\begin{tabular}{|c|c|c|}
\hline
          & The length of each transmitted symbol vector & The average numbers of transmitted symbols per sentence \\ \hline
Proposed  & 32                                    & 1120                                                 \\ \hline
DeepSC-SR & 20                                    & 7143                                                    \\ \hline
SE-DeepSC & 128                                   & 2225                                                    \\ \hline
DeepSC-S  & 16384                                     &  655360                                                \\ \hline
\end{tabular}
\end{table}

%这个表格留到最后。可以控制文章的长度。
\begin{table*}[htbp]
  \centering
  \caption{Examples of input speech signals and their transcriptions before and after the redundancy removal module.}
  \label{tablelong}
  \begin{tabular}{|c|p{0.8\linewidth}|}
  \hline
    Example 1  & The number of speech spectrum frames: 497 \\
    & Transcription: (Length:36) $BOS$ RANDAL HE SAID YOU KNOW WHERE SYDNEY IS $EOS$ $\times$ 27  \\ saved: 77.8\% %$EOS$ $EOS$ $EOS$ $EOS$ $EOS$ $EOS$ $EOS$ $EOS$ $EOS$ $EOS$ $EOS$ $EOS$ $EOS$ $EOS$ $EOS$ $EOS$ $EOS$ $EOS$ $EOS$ $EOS$ $EOS$ $EOS$ $EOS$ $EOS$ $EOS$ $EOS$ \\ saved: 85.3\%
& Transcription after the redundancy removal module: (Length:8)RANDAL HE SAID YOU KNOW WHERE SYDNEY IS\\ \hline
Example 2 & The number of speech spectrum frames: 9845 \\
& Transcription: (Length:96) $BOS$ I HAVE DRAWN UP A LIST OF ALL THE PEOPLE WHO OUGHT TO GIVE US A PRESENT AND I SHALL TELL THEM WHAT THEY OUGHT TO GIVE IT WON'T BE MY FAULT IF I DON'T GET IT $EOS$ $EOS$ $EOS$ $EOS$ $EOS$ $EOS$ OF ALL THE PEOPLE WHO OUGHT TO GIVE US A PRESENT AND I SHALL TELL THEM WHAT THEY OUGHT TO GIVE IT WON'T BE MY FAULT IF I DON'T GET IT $EOS$ $EOS$ $EOS$ $EOS$ $EOS$ OF ALL THE PEOPLE WHO OUGHT TO GIVE US A PRESENT AND I SHALL TELL THEM WHAT THEY OUGHT TO GIVE IT WON'T BE MY FAULT IF \\ saved: 61.5\%
& Transcription after the redundancy removal module: (Length:37) I HAVE DRAWN UP A LIST OF ALL THE PEOPLE WHO OUGHT TO GIVE US A PRESENT AND I SHALL TELL THEM WHAT THEY OUGHT TO GIVE IT WON'T BE MY FAULT IF I DON'T GET IT \\ \hline
%Sample 4 & Speech Length: 14650 \\
%& Redundant: (Length:143) HE $UNK$ AT AN EARLY AGED $UNK$ BETWEEN EIGHTEEN FORTY FIVE AND EIGHTEEN FORTY NINE A GREAT NUMBER OF POEMS BY THE ITALIANS CONTEMPORARY WITH  $UNK$ OR PROCEEDING HIM AND AMONG OTHER THINGS HE MADE AVERSION OF THE WHOLE $UNK$ $UNK$ AND VERSE $EOS$ $EOS$ $EOS$ $EOS$ $EOS$ A GREAT NUMBER OF POEMS BY THE ITALIANS CONTEMPORARY WITH $UNK$ OR PROCEEDING HIM AND AMONG OTHER THINGS HE MADE AVERSION OF THE WHOLE $UNK$ $EOS$ AND VERSE $EOS$ $EOS$ $EOS$ $EOS$ $EOS$ A GREAT NUMBER OF POEMS BY THE ITALIANS CONTEMPORARY WITH $UNK$ OR PROCEEDING HIM AND AMONG OTHER THINGS HE MADE AVERSION OF THE WHOLE $UNK$ $EOS$ AND VERSE $EOS$ $EOS$ $EOS$ $EOS$ $EOS$ A GREAT NUMBER OF POEMS BY THE ITALIANS CONTEMPORARY WITH $UNK$ OR PROCEEDING HIM AND AMONG OTHER THINGS HE MADE AVERSION OF THE WHOLE $UNK$ $EOS$ AND VERSE $EOS$ \\ saved:73.4\%
%& Non-redundant: (Length:38) HE AT AN EARLY AGED BETWEEN EIGHTEEN FORTY FIVE AND EIGHTEEN FORTY NINE A GREAT NUMBER OF POEMS BY THE ITALIANS CONTEMPORARY WITH OR PROCEEDING HIM AND AMONG OTHER THINGS HE MADE AVERSION OF THE WHOLE AND VERSE\\\hline
\end{tabular}
\end{table*}

%69405509(58822747$+$10582762) 
% 349.2MB(225MB$+$62.1MB)

\subsection{Transmission Efficiency}

We present the length of the transmitted symbol vector and the average number of the transmitted symbols per sentence on the same testing dataset of different approaches in Table. \ref{table_number}. The length of each transmitted symbol vector is half of the dimension of each channel encoder output since two elements are combined into a complex value to transmit. We can see from this table that although our proposed model has slightly longer symbol vectors, i.e., a larger dimension of the output of the channel encoder, the average number of transmitted symbols per sentence by the proposed model is 16\% of that by DeepSC-SR, significantly improving the transmission efficiency while achieving about 10\% WER reduction as shown in Fig. 6. The reason may be that DeepSC-SR encodes every single speech spectrum frame into the same amount of transmitted symbols, while the proposed method ignores the redundant content, and only sends the text-related semantic information.

Two examples of the speech signals and their corresponding transcriptions as well as the transcription after redundancy removal are shown in Table. \ref{tablelong}. Both of the examples show that the number of the speech spectrum frames is much larger than the length of the transcription, which implies that many of the frames are semantically irrelevant. We also observe that the redundancy removal module significantly reduces the length of the transcription while preserving the semantic meaning. 

%We can see from this table that although our proposed model has much longer symbol vectors, i.e., a larger dimension of the output of the channel encoder, the average number of transmitted symbols per sentence by the proposed model is still much smaller than that by DeepSC-SR. The reason may be that DeepSC-SR encodes every single speech spectrum frame into the same amount of transmitted symbols, while the proposed method ignores the redundant content, and only sends the text-related semantic information.

%we also present the relation between the length of the transmitted symbol vector and the system accuracy under same channel conditions in Fig. \ref{bangdwith.pdf}.

\begin{figure}[tbp]
\includegraphics[width=0.5\textwidth]{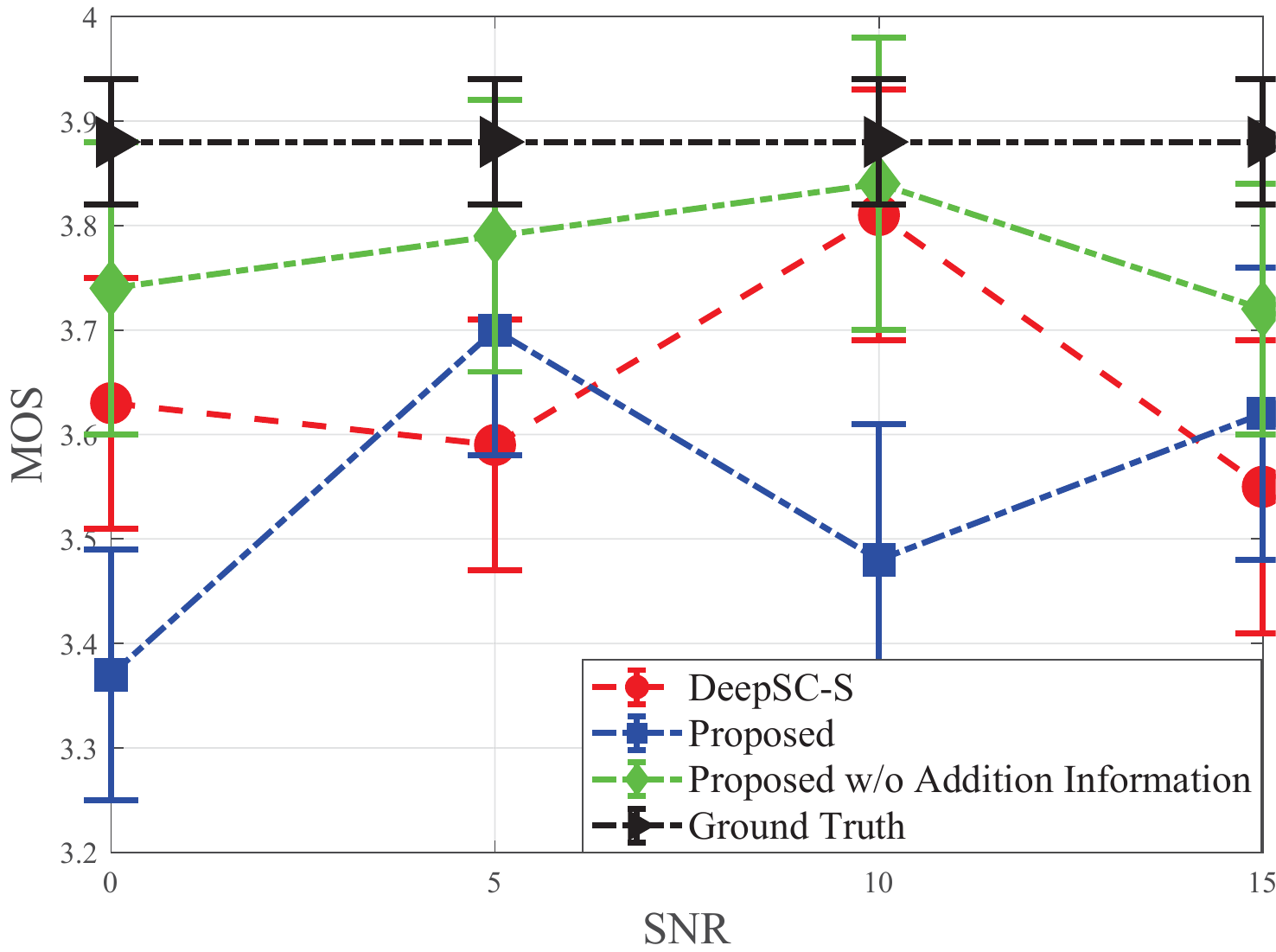}
\centering % central position
\caption{MOS versus SNR under AWGN channel for different speech to speech transmission approaches.}  % figure title
\label{MOS}  
\end{figure}

\begin{figure}[tbp]
\includegraphics[width=0.5\textwidth]{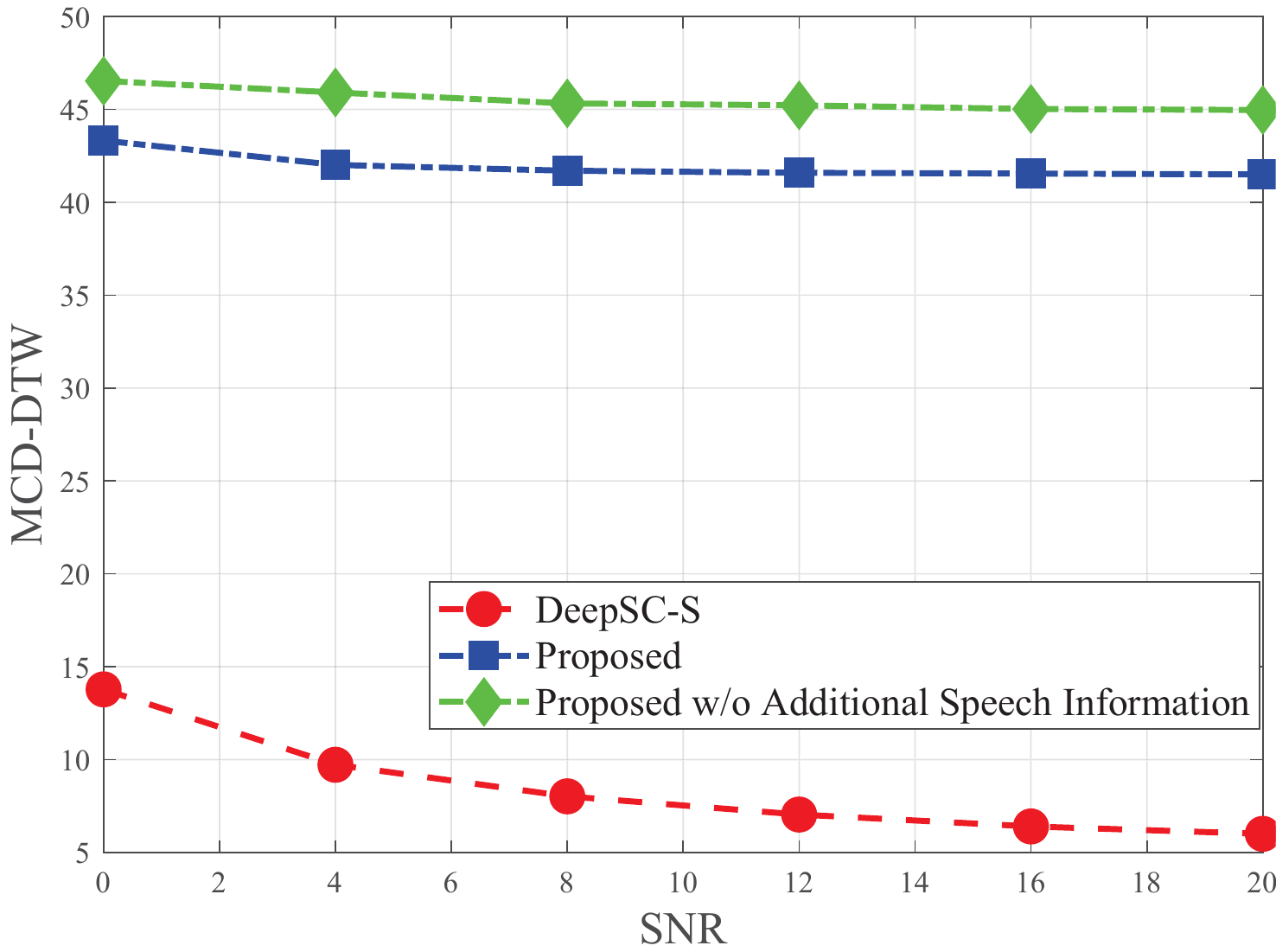}
\centering % central position
\caption{MCD versus SNR under AWGN channel for different speech to speech transmission approaches.}  % figure title
\label{MCD}  
\end{figure}

%\begin{figure}[tbp]
%\centering % central position
  % figure title
%\includegraphics[width=0.4\textwidth]{figure/spectrumnew.pdf}
%\caption{The spectrum of the recovered speech from our proposed system. }
%\label{spectrum_reslut}  
%\end{figure}

\begin{figure}
\centering % central position
  % figure title
  \subfigure{
\begin{minipage}{0.45\textwidth}
    \centering
    \includegraphics[width=\textwidth]{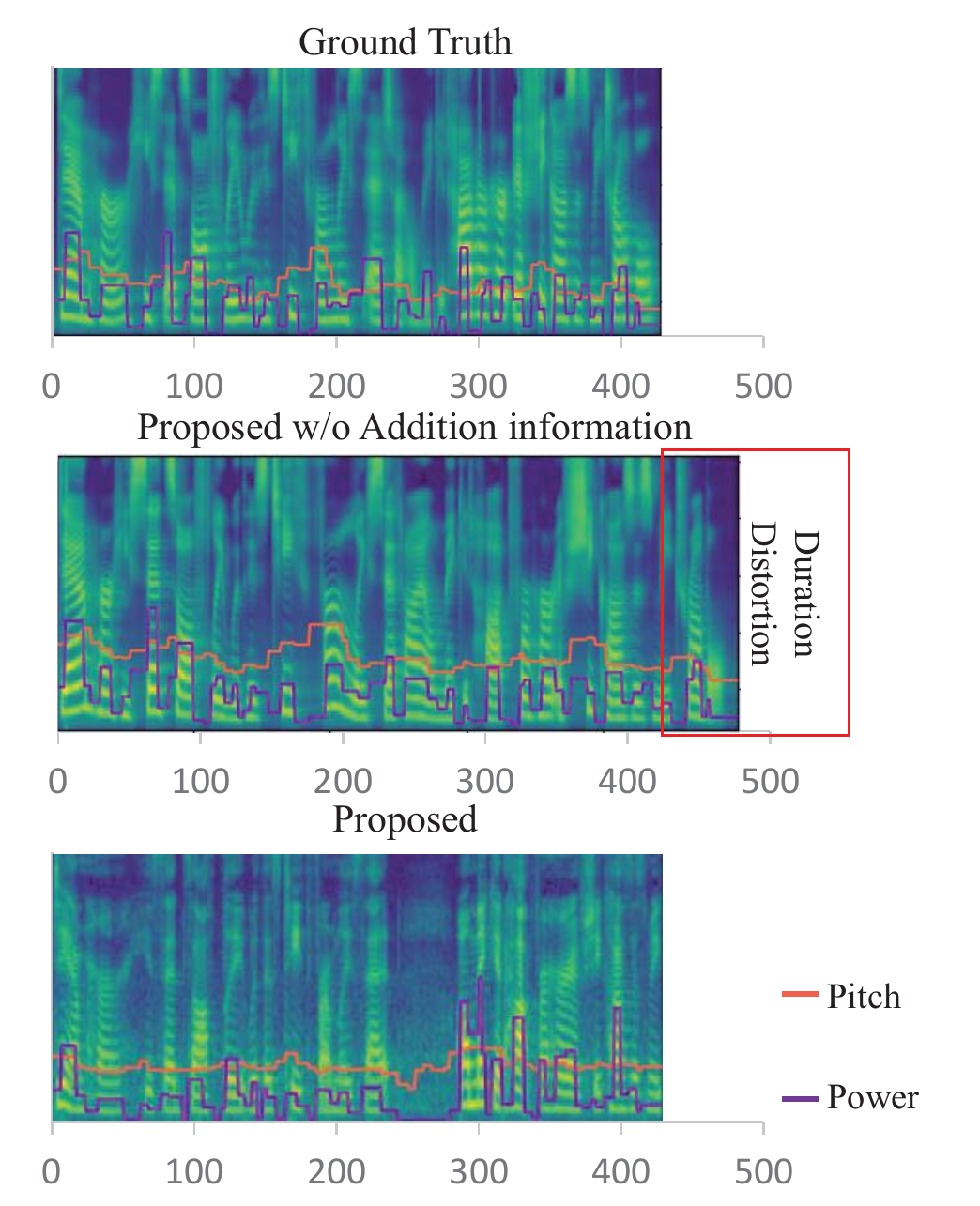}
    %\subcaption{AWGN channels}
\end{minipage}}
%\qquard
\subfigure{
\begin{minipage}{0.45\textwidth}
    \centering
    \includegraphics[width=\textwidth]{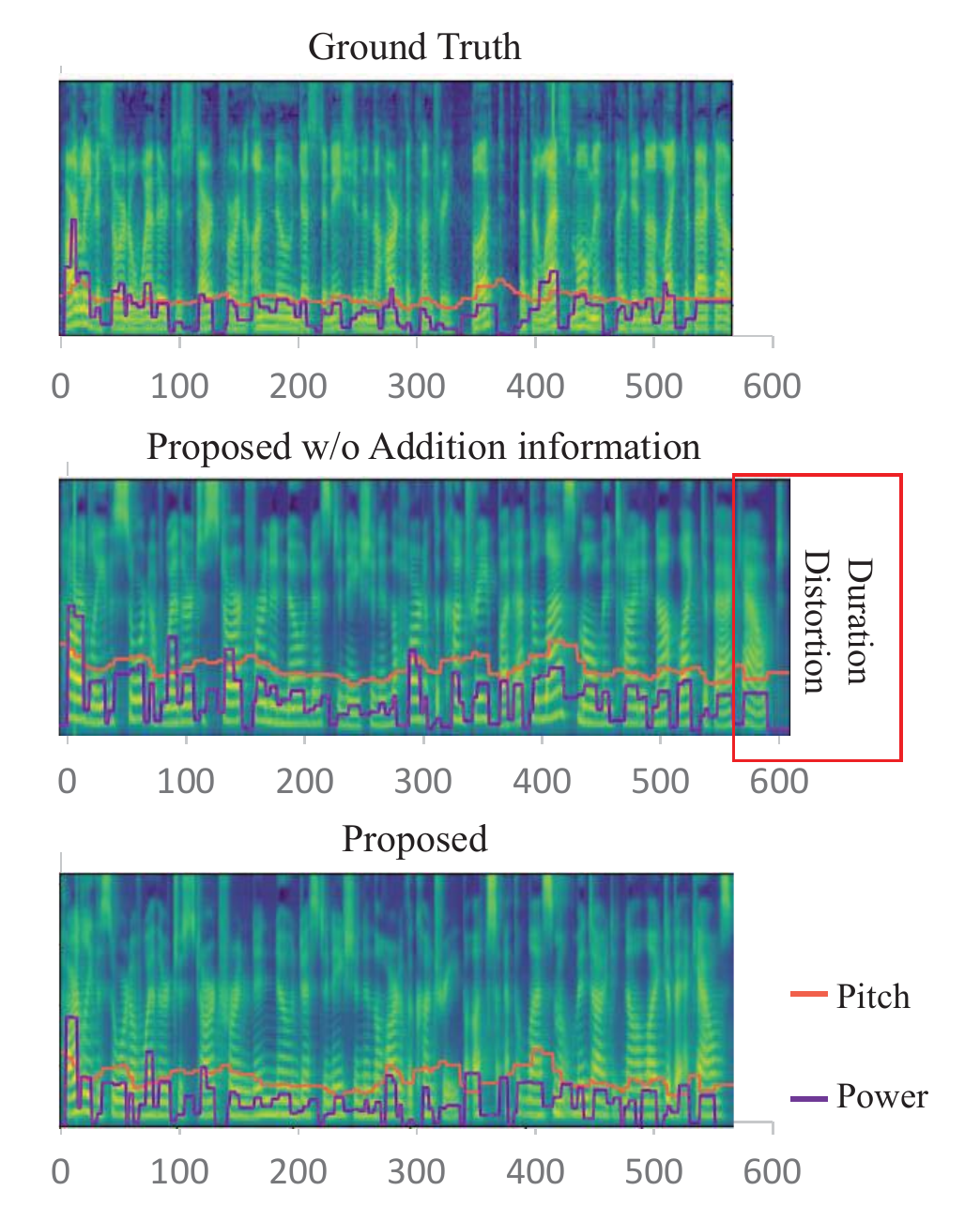}
    %\subcaption{AWGN channels}
\end{minipage}}
\caption{The comparison between the reconstructed spectrums and the original spectrums.}
\label{spectrum_reslut}  
\end{figure}

\subsection{Performance comparison of the speech to speech transmission}
We then compare the quality of the recovered speech by the proposed approach and DeepSC-S in terms of MOS and MCD, as shown in Fig. \ref{MOS} and Fig. \ref{MCD}, respectively. The MOS scores presented in Fig. \ref{MOS} are obtained by 20 native speakers listening to 40 generated speech samples by each approach and rating the scores according to Table \ref{MOStable}. As it can be observed from Fig. \ref{MOS}, our system achieves comparable results with DeepSC-S. However, surprisingly, the MOS score of our proposed method without additional speech information is better than with additional information, and very close to the ground truth results. The reason may be that the MOS score measures the naturalness of the generated speech instead of the closeness to the original speech. And the adopted pre-trained GAN vocoder is powerful to convert the spectrum into natural speech. However, we can observe significant performance gain achieved by the additional information provided in Fig. \ref{MCD} in terms of MCD, which measures the distance between the recovered speech spectrums and the original ones. This implies the additional speech information helps the recovery of the original speech spectrums. We note the DeepSC-S results in much lower MCD compared to the proposed approach. This is due to the fact that DeepSC-S views speech spectrums as images, and aims for exact recovery of the spectrums including the noises in the spectrum. However, our approach focuses on the successful transmission of semantic information while eliminating the semantic irrelevant factors in the speech spectrum, such as environment noises. We also present the predicted speech spectrums by the proposed approach and the original spectrums in Fig. \ref{spectrum_reslut}. We can see that the predicted spectrum without the additional speech information provided is longer than the original speech, and the pitch and power less similar to the ground truth compared to that by the proposed approach with additional information provided. We provide the recovered audio samples at \url{https://txhan.github.io/speech-to-text/}.

In terms of transmission efficiency, we note that the additional information only needs to be sent once for each utterances, and is of a small dimension of $64$, which can be neglected compared to the amount of the semantic information sent over the channel. We can see from the Table \ref{table_number} that the proposed approach for the speech to speech transmission sends $1120$ transmitted symbols per sentence over the channel, less than 0.2\% of that by the existing method, DeepSC-S, which results in $655360$ transmitted symbols per sentence.

\section{Conclusion}
In this paper, we proposed a semantic-oriented communication system for speech transmission, which transmits only the semantic-relevant information over the channel for the speech recognition task, and a compact additional set of semantic-irrelevant information for the speech reconstruction task to improve transmission efficiency. In particular, for the \textit{speech to text} transmission, we employed an attention-based soft alignment module and a redundancy removal module to extract only the text-related semantic features while dropping semantically redundant content, which reduces the 90\% of the latent semantic features by the semantic encoder as revealed by the numerical results. We also introduced a beam search semantic decoder to find the most possible transcription exploiting the long time dependency in the input sequences, and a semantic correction module based on a pretrained language model to further correct the predicted transcription with semantic knowledge learnt from a large and comprehensive text dataset. For the speech to speech transmission, we further included a CTC alignment module at the receiver side to extract additional information from the original speech that helps the recovery of the speech including the duration, pitch and power information of each phoneme, and a speech reconstructor at the receiver which leverages a text to speech decoder and a pretrained GAN to reconstruct the speech signals by combining the received semantic information and the additional speech-related information. We also introduced a two-stage training scheme, which trains different parts of the proposed model at each stage, and, thus, speeds up the training of the model as revealed by experiments. Simulation results demonstrate that our proposed method outperforms our previous work and the existing methods in terms of the accuracy of the predicted text for the speech to text transmission, and achieves comparable results with the existing approach in terms of the quality of the recovered speech signals for the speech to speech transmission. It is worth mentioning that the proposed approach significantly improves the transmission efficiency over the existing methods with only 16\% of the number of transmitted symbols per sentence required the state-of-the-art approach for the speech to text transmission, and only transmits shockingly 0.2\% of the amount of transmitted symbols by the existing method for the speech to speech transmission.

\bibliographystyle{IEEEtran}
\bibliography{main.bib}

\begin{thebibliography}{10}
\providecommand{\url}[1]{#1}
\csname url@samestyle\endcsname
\providecommand{\newblock}{\relax}
\providecommand{\bibinfo}[2]{#2}
\providecommand{\BIBentrySTDinterwordspacing}{\spaceskip=0pt\relax}
\providecommand{\BIBentryALTinterwordstretchfactor}{4}
\providecommand{\BIBentryALTinterwordspacing}{\spaceskip=\fontdimen2\font plus
\BIBentryALTinterwordstretchfactor\fontdimen3\font minus
  \fontdimen4\font\relax}
\providecommand{\BIBforeignlanguage}[2]{{%
\expandafter\ifx\csname l@#1\endcsname\relax
\typeout{** WARNING: IEEEtran.bst: No hyphenation pattern has been}%
\typeout{** loaded for the language `#1'. Using the pattern for}%
\typeout{** the default language instead.}%
\else
\language=\csname l@#1\endcsname
\fi
#2}}
\providecommand{\BIBdecl}{\relax}
\BIBdecl

\bibitem{qin2021semantic_survey}
Z.~Qin, X.~Tao, J.~Lu, and G.~Y. Li, ``Semantic communications: Principles and
  challenges,'' \emph{arXiv preprint arXiv:2201.01389}, 2021.

\bibitem{qin1}
H.~Xie, Z.~Qin, G.~Y. Li, and B.-H. Juang, ``Deep learning enabled semantic
  communication systems,'' \emph{IEEE Trans. Signal Process.}, pp. 2663--2675,
  Apr. 2021.

\bibitem{shannon1949mathematical}
C.~E. Shannon, \emph{The mathematical theory of communication, by CE Shannon
  (and recent contributions to the mathematical theory of communication), W.
  Weaver}.\hskip 1em plus 0.5em minus 0.4em\relax University of illinois Press
  Champaign, IL, USA, 1949.

\bibitem{carnap1952outline}
R.~Carnap, Y.~Bar-Hillel \emph{et~al.}, \emph{An Outline of A Theory of
  Semantic Information}.\hskip 1em plus 0.5em minus 0.4em\relax RLE Technical
  Reports 247, Research Laboratory of Electronics, Massachusetts Institute of
  Technology., Cambridge MA, Oct. 1952.

\bibitem{floridi2004outline}
L.~Floridi, ``Outline of a theory of strongly semantic information,''
  \emph{Minds and machines}, vol.~14, no.~2, pp. 197--221, 2004.

\bibitem{bao2011towards}
J.~Bao, P.~Basu, M.~Dean, C.~Partridge, A.~Swami, W.~Leland, and J.~A. Hendler,
  ``Towards a theory of semantic communication,'' in \emph{2011 IEEE Network
  Science Workshop}.\hskip 1em plus 0.5em minus 0.4em\relax IEEE, 2011, pp.
  110--117.

\bibitem{basu2014preserving}
P.~Basu, J.~Bao, M.~Dean, and J.~Hendler, ``Preserving quality of information
  by using semantic relationships,'' \emph{Pervasive and Mobile Computing},
  vol.~11, pp. 188--202, 2014.

\bibitem{bourtsoulatze2019deep}
E.~Bourtsoulatze, D.~B. Kurka, and D.~G{\"u}nd{\"u}z, ``Deep joint
  source-channel coding for wireless image transmission,'' \emph{IEEE Trans.
  Cogn. Commun. Netw.}, vol.~5, no.~3, pp. 567--579, Sept. 2019.

\bibitem{kurka2021bandwidth}
D.~B. Kurka and D.~G{\"u}nd{\"u}z, ``Bandwidth-agile image transmission with
  deep joint source-channel coding,'' \emph{IEEE Transactions on Wireless
  Communications}, vol.~20, no.~12, pp. 8081--8095, 2021.

\bibitem{zhang2022wireless}
Z.~Zhang, Q.~Yang, S.~He, M.~Sun, and J.~Chen, ``Wireless transmission of
  images with the assistance of multi-level semantic information,'' \emph{arXiv
  preprint arXiv:2202.04754}, 2022.

\bibitem{farsad2018deep}
N.~Farsad, M.~Rao, and A.~Goldsmith, ``Deep learning for joint source-channel
  coding of text,'' in \emph{2018 IEEE international conference on acoustics,
  speech and signal processing (ICASSP)}.\hskip 1em plus 0.5em minus
  0.4em\relax IEEE, 2018, pp. 2326--2330.

\bibitem{rao2018variable}
M.~Rao, N.~Farsad, and A.~Goldsmith, ``Variable length joint source-channel
  coding of text using deep neural networks,'' in \emph{2018 IEEE 19th
  international workshop on signal processing advances in wireless
  communications (SPAWC)}.\hskip 1em plus 0.5em minus 0.4em\relax IEEE, 2018,
  pp. 1--5.

\bibitem{xie2020lite}
H.~Xie and Z.~Qin, ``A lite distributed semantic communication system for
  {I}nternet of {T}hings,'' \emph{IEEE J. Sel. Areas Commun.}, vol.~39, no.~1,
  pp. 142--153, Jan. 2021.

\bibitem{xie2021task}
H.~Xie, Z.~Qin, X.~Tao, and K.~B. Letaief, ``Task-oriented multi-user semantic
  communications,'' \emph{arXiv preprint arXiv:2112.10255}, 2021.

\bibitem{xie2021task1}
H.~Xie, Z.~Qin, and G.~Y. Li, ``Task-oriented multi-user semantic
  communications for vqa task,'' \emph{IEEE Wireless Communications Letters},
  2021.

\bibitem{tung2021deepwive}
T.-Y. Tung and D.~G{\"u}nd{\"u}z, ``Deepwive: Deep-learning-aided wireless
  video transmission,'' \emph{arXiv preprint arXiv:2111.13034}, 2021.

\bibitem{shao2021learning}
J.~Shao, Y.~Mao, and J.~Zhang, ``Learning task-oriented communication for edge
  inference: An information bottleneck approach,'' \emph{IEEE Journal on
  Selected Areas in Communications}, vol.~40, no.~1, pp. 197--211, 2021.

\bibitem{Weng2101:Semantic}
Z.~Weng and Z.~Qin, ``Semantic communication systems for speech transmission,''
  \emph{IEEE J. Sel. Areas Commun.}, Apr. 2021.

\bibitem{tong2021federated}
H.~Tong, Z.~Yang, S.~Wang, Y.~Hu, O.~Semiari, W.~Saad, and C.~Yin, ``Federated
  learning for audio semantic communication,'' \emph{Frontiers in
  Communications and Networks}, vol.~2, 2021.

\bibitem{weng2021semantic}
Z.~Weng, Z.~Qin, and G.~Y. Li, ``Semantic communications for speech
  recognition,'' \emph{arXiv preprint arXiv:2107.11190}, 2021.

\bibitem{graves2006connectionist}
A.~Graves, S.~Fern{\'a}ndez, F.~Gomez, and J.~Schmidhuber, ``Connectionist
  temporal classification: Labelling unsegmented sequence data with recurrent
  neural networks,'' in \emph{Proc. 23rd Int. Conf. Mach. Learning (ICML)},
  Pittsburgh, USA, Jun. 2006, pp. 369--376.

\bibitem{hermansky1990perceptual}
H.~Hermansky, ``Perceptual linear predictive (plp) analysis of speech,''
  \emph{the Journal of the Acoustical Society of America}, vol.~87, no.~4, pp.
  1738--1752, 1990.

\bibitem{bahdanau2015neural}
D.~Bahdanau, K.~H. Cho, and Y.~Bengio, ``Neural machine translation by jointly
  learning to align and translate,'' in \emph{3rd International Conference on
  Learning Representations, ICLR 2015}, 2015.

\bibitem{bazzi2002modelling}
I.~Bazzi, ``Modelling out-of-vocabulary words for robust speech recognition,''
  Ph.D. dissertation, Massachusetts Institute of Technology, 2002.

\bibitem{han2022semantic}
T.~Han, Q.~Yang, Z.~Shi, S.~He, and Z.~Zhang, ``Semantic-aware speech to text
  transmission with redundancy removal,'' \emph{arXiv preprint
  arXiv:2202.03211}, 2022.

\bibitem{kudo2018sentencepiece}
T.~Kudo and J.~Richardson, ``Sentencepiece: A simple and language independent
  subword tokenizer and detokenizer for neural text processing,'' in
  \emph{Proceedings of the 2018 Conference on Empirical Methods in Natural
  Language Processing: System Demonstrations}, 2018, pp. 66--71.

\bibitem{ren2020fastspeech}
Y.~Ren, C.~Hu, X.~Tan, T.~Qin, S.~Zhao, Z.~Zhao, and T.-Y. Liu, ``Fastspeech 2:
  Fast and high-quality end-to-end text to speech,'' in \emph{International
  Conference on Learning Representations}, 2020.

\bibitem{klakow2002testing}
D.~Klakow and J.~Peters, ``Testing the correlation of word error rate and
  perplexity,'' \emph{Speech Communication}, vol.~38, no. 1-2, pp. 19--28,
  2002.

\bibitem{devlin2018bert}
J.~D. M.-W.~C. Kenton and L.~K. Toutanova, ``Bert: Pre-training of deep
  bidirectional transformers for language understanding,'' in \emph{Proceedings
  of NAACL-HLT}, 2019, pp. 4171--4186.

\bibitem{kubichek1993mel}
R.~Kubichek, ``Mel-cepstral distance measure for objective speech quality
  assessment,'' in \emph{Proceedings of IEEE pacific rim conference on
  communications computers and signal processing}, vol.~1.\hskip 1em plus 0.5em
  minus 0.4em\relax IEEE, 1993, pp. 125--128.

\bibitem{ribeiro2011crowdmos}
F.~Ribeiro, D.~Flor{\^e}ncio, C.~Zhang, and M.~Seltzer, ``Crowdmos: An approach
  for crowdsourcing mean opinion score studies,'' in \emph{2011 IEEE
  international conference on acoustics, speech and signal processing
  (ICASSP)}.\hskip 1em plus 0.5em minus 0.4em\relax IEEE, 2011, pp. 2416--2419.

\bibitem{vgg}
K.~Simonyan and A.~Zisserman, ``Very deep convolutional networks for
  large-scale image recognition,'' in \emph{International Conference on
  Learning Representations}, May 2015.

\bibitem{chorowski2015attention}
J.~K. Chorowski, D.~Bahdanau, D.~Serdyuk, K.~Cho, and Y.~Bengio,
  ``Attention-based models for speech recognition,'' \emph{Advances in neural
  information processing systems}, vol.~28, 2015.

\bibitem{forney1973viterbi}
G.~D. Forney, ``The viterbi algorithm,'' \emph{Proceedings of the IEEE},
  vol.~61, no.~3, pp. 268--278, 1973.

\bibitem{g2pE2019}
J.~Park, Kyubyong~Kim, ``g2pe,'' \url{https://github.com/Kyubyong/g2p}, 2019.

\bibitem{kong2020hifi}
J.~Kong, J.~Kim, and J.~Bae, ``Hifi-gan: Generative adversarial networks for
  efficient and high fidelity speech synthesis,'' \emph{Advances in Neural
  Information Processing Systems}, vol.~33, pp. 17\,022--17\,033, 2020.

\bibitem{ko2015audio}
T.~Ko, V.~Peddinti, D.~Povey, and S.~Khudanpur, ``Audio augmentation for speech
  recognition,'' in \emph{Sixteenth annual conference of the international
  speech communication association}, 2015.

\bibitem{ko2017study}
T.~Ko, V.~Peddinti, D.~Povey, M.~L. Seltzer, and S.~Khudanpur, ``A study on
  data augmentation of reverberant speech for robust speech recognition,'' in
  \emph{2017 IEEE International Conference on Acoustics, Speech and Signal
  Processing (ICASSP)}.\hskip 1em plus 0.5em minus 0.4em\relax IEEE, 2017, pp.
  5220--5224.

\bibitem{williams1989learning}
R.~J. Williams and D.~Zipser, ``A learning algorithm for continually running
  fully recurrent neural networks,'' \emph{Neural computation}, vol.~1, no.~2,
  pp. 270--280, 1989.

\bibitem{panayotov2015librispeech}
V.~Panayotov, G.~Chen, D.~Povey, and S.~Khudanpur, ``Librispeech: an asr corpus
  based on public domain audio books,'' in \emph{2015 IEEE international
  conference on acoustics, speech and signal processing (ICASSP)}.\hskip 1em
  plus 0.5em minus 0.4em\relax IEEE, 2015, pp. 5206--5210.

\end{thebibliography}

\end{document}